\documentclass[aps,floats]{revtex4}
\usepackage{amsmath,amssymb}
\usepackage{graphicx,epsfig}
\usepackage[greek,english]{babel}
\usepackage{bbold}
\usepackage{braket}

\begin{document}
\bibliographystyle {plain}

\pdfoutput=1
\def\oppropto{\mathop{\propto}} 
\def\opsimeq{\mathop{\simeq}}
\def\opoverderline{\mathop{\overline}}
\def\operarrow{\mathop{\longrightarrow}}
\def\opsim{\mathop{\sim}}

\def\fig#1#2{\includegraphics[height=#1]{#2}}
\def\figx#1#2{\includegraphics[width=#1]{#2}}


\title{ Properties of the simplest inhomogeneous and homogeneous Tree-Tensor-States \\
for long-ranged quantum spin chains with or without disorder 
}


\author{ C\'ecile Monthus }
 \affiliation{Institut de Physique Th\'{e}orique, 
Universit\'e Paris Saclay, CNRS, CEA,
91191 Gif-sur-Yvette, France}

\begin{abstract}
The simplest Tree-Tensor-States (TTS) respecting the Parity and the Time-Reversal symmetries are studied in order to describe the ground states of long-ranged quantum spin chains with or without disorder. Explicit formulas are given for the one-point and two-point reduced density matrices that allow to compute any one-spin and two-spin observable. For Hamiltonians containing only one-body and two-body contributions, the energy of the TTS can be then evaluated and minimized in order to obtain the optimal parameters of the TTS. This variational optimization of the TTS parameters is compared with the traditional block-spin renormalization procedure based on the diagonalization of some intra-block renormalized Hamiltonian.

\end{abstract}

\maketitle


 \section{ Introduction }

The entanglement between the different regions of many-body quantum systems
(see the reviews \cite{amico08,horo,calabrese09,qi,laflorencie16,chiara} and references therein)
has emerged as an essential physical property that should be taken into account in their descriptions.
In the field of Tensor Networks
(see the reviews \cite{wolf,ver,cirac,vidal_intro,ulrich2011,phd-evenbly,mera-review,hauru,orus14a,orus14b,orus19}
 and references therein), the ground-state wavefunction is decomposed 
into elementary small tensors that can be assembled 
in various ways in order to adapt to the geometry, to the symmetries, and to the entanglement properties of
the problem under focus.
In particular, various previous real-space renormalization procedures 
for the ground states of quantum spin chains
have been reinterpreted and possibly improved within this new perspective.
For instance, the Density-Matrix-Renormalization-Group (DMRG)  \cite{white1,white2,ulrich2005}
was reformulated  as a variational problem based on 
Matrix-Products-States (MPS) that are well adapted to describe non-critical states
displaying area-law entanglement.
The traditional block-spin renormalization for critical points corresponds to scale-invariant Tree-Tensor-States (TTS),
and has been improved via the multi-scale-entanglement-renormalization-ansatz (MERA)
\cite{vidal_mera,evenbly_mera}, where 'disentanglers' between blocks are introduced besides the block-coarse-graining isometries already present in Tree-Tensor-States. 
Finally in the field of disordered spin chains, 
the Strong Disorder Renormalization Group (SDRG) approach introduced by Daniel Fisher \cite{danielrtfic,danielantiferro,danielreview} (see the reviews \cite{strong_review1,strong_review2} for further references)
has been reformulated either as a Matrix-Product-Operator-Renormalization or as a self-assembling Tree-Tensor-Network,
and various improvements have been proposed 
\cite{sigrist,romer,lin1,gold,chatelain,lin2}.

However, even in the second example
 where the 'old' block-spin renormalization procedure and the 'new' Tree-Tensor-State variational approach
share the same entanglement architecture, the precise choice of the elementary isometries remains different.
Indeed in the traditional block-spin renormalization, the isometries are determined via the diagonalization of some 
'intra-block' Hamiltonian involving a few renormalized spins, so that one can usually obtain 
explicit RG flows for the parameters of the renormalized Hamiltonian.
The two main criticisms levelled against this procedure can be summarized as follows :
(i) at the level of principles, the choice of the ground state of the 'intra-block Hamiltonian' 
does not take at all into account the 'environment' of the neighboring blocks;
(ii) in practice, there is usually some arbitrariness in the decomposition of the Hamiltonian into
the 'intra-block' and the 'extra-block' contributions that can lead to completely different outputs, 
so that the quality of the results strongly depends on the cleverness of the choice of the intra-block Hamiltonian.
In the Tree-Tensor perspective, one considers instead the whole ground-state wavefunction 
 as a variational tree-tensor-state involving isometries, and the optimization of each isometry is based 
on the minimization of the total energy of the Tree-Tensor-State.
At the level of principles, the theoretical advantage is clearly
that the output corresponds to the optimal Tree-Tensor-State,
i.e. to the best renormalization procedure within the class of all renormalization procedures of a given dimension.
In practice, the drawback is that this global optimization is more complicated and can usually be done only numerically,
unless the isometries are completely fixed by the very strong quantum symmetries of the model \cite{c_xxzq}.

In the present paper, the goal is to analyze the explicit properties of the simplest Tree-Tensor-States of
the smallest bond dimension $D=2$ in the context of long-ranged quantum spin chains with or without disorder,
in order to analyze more precisely the improvement given by the global optimization of the isometries 
with respect to the traditional block-spin procedure.

The paper is organized as follows.
In section \ref{sec_LR}, we introduce the notations for long-ranged 
quantum spin chains with Parity and Time-Reversal symmetries. 
In section \ref{sec_treetensor}, 
the elementary RG step for a block of two spins is parametrized is terms of two angles.
In section \ref{sec_descending}, 
we describe the simplest inhomogeneous Tree-Tensor-States;
the explicit forms of the one-point and two-point reduced density matrices
are derived in order to analyze the structure of magnetizations and two-points correlations.
In section \ref{sec_energy}, we focus on the energy of the Tree-Tensor-State 
for disordered long-ranged spin Chains
in order to discuss the optimization with respect to the Tree-Tensor-States parameters.
In section \ref{sec_pure}, we turn to the case of pure long-ranged spin chains
in order to take into account the supplementary symmetries in the Tree-Tensor-States.
Finally in section \ref{sec_criti}, we study the properties of the scale-invariant Tree-Tensor-States
for the critical points of pure models.
Our conclusions are summarized in section \ref{sec_conclusion}.
The appendix \ref{app_diago} contains the traditional block-spin determination of the parameters of the Tree-Tensor-State, in order to compare with the variational optimization discussed in the text.


\section{ Long-ranged spin chains with Parity and Time-Reversal Symmetries }

\label{sec_LR}

Within the Tensor-Network perspective, the symmetries play an essential role 
in order to restrict the form of the possible isometries.
In the present paper, we focus on quantum spin chains with Parity and Time-Reversal Symmetries.

\subsection{ Parity and Time-Reversal operators }

For a quantum spin chain of $N$ spins described by Pauli matrices $\sigma_n^{a=0,x,y,z} $, 
the Parity operator
\begin{eqnarray}
{\cal P}= \prod_{n=1}^N \sigma_n^z
\label{parityop}
\end{eqnarray}
and the Time-Reversal operator ${\cal T}$ whose action can be defined via
\begin{eqnarray}
{\cal T}  i {\cal T}^{-1} && =  -i
\nonumber \\
{\cal T}  \sigma_n^x {\cal T}^{-1} && =   \sigma_n^x
\nonumber \\
{\cal T}   \sigma_n^y {\cal T}^{-1} &&=  -  \sigma_n^y
\nonumber \\
{\cal T}   \sigma_n^z {\cal T}^{-1} && =  \sigma_n^z
\label{timeop}
\end{eqnarray}
are among the most important possible symmetries.


\subsection{Hamiltonians containing only one-body and two-body terms respecting the Parity and Time-Reversal}

\label{sec_lrsym}

The Hamiltonians commuting with the Parity and Time-Reversal operators 
and containing only one-body and two-body terms
can be parametrized in terms of fields $h_n$ and in terms of couplings ${\cal J}_{n,n'}^{a=x,y,z}$ as
\begin{eqnarray}
{\cal H}_N= - \sum_{n=1}^N h_n \sigma_n^z
 - \sum_{1 \leq n<n' \leq N } (  {\cal J}_{n,n'}^x\sigma_n^x \sigma_{n'}^x
+ {\cal J}_{n,n'}^y \sigma_n^y \sigma_{n'}^y
+{\cal J}_{n,n'}^z\sigma_n^z \sigma_{n'}^z )
\label{hpt}
\end{eqnarray}
Then one needs to choose whether the fields $ h_n$ are uniform or random,
whether the couplings ${\cal J}_{n,n'}^{a=x,y,z} $ are short-ranged or long-ranged, with or without disorder.

We should stress here that the Parity and the Time-Reversal 
are the only symmetries that will be taken into account in the present paper,
while the models displaying further symmetries like magnetization conservation 
(corresponding to the identity between $x$ and $y$ couplings $ {\cal J}_{n,n'}^x={\cal J}_{n,n'}^y$) 
or $SU(2)$ invariance (corresponding to the identity between $x$, $y$ and $z$ couplings $ {\cal J}_{n,n'}^x={\cal J}_{n,n'}^y={\cal J}_{n,n'}^z$)
would require other isometries in order to take into account these stronger symmetry properties.


\subsection{Long-ranged quantum spin chains with power-law couplings}

In the present paper, we will focus on the cases
where the couplings $ {\cal J}_{n,n'}^{a=x,y,z}$ are long-ranged
with some power-law dependence with respect to the distance $r(n,n')=\vert n-n' \vert $ between the two sites $n$ and $n'$
\begin{eqnarray}
{\cal J}_{n,n'}^a = \frac{J^a_{n,n'} }{\left[ r(n,n')  \right]^{1+\omega_a } } = \frac{J^a_{n,n'}}{ \vert n-n' \vert^{1+\omega_a } } 
\label{jpower}
\end{eqnarray}
where the amplitudes $J^a_{n,n'} $ are of order unity,
while the exponents $\omega_a$ governing the decays with the distance 
are positive $\omega_a>0$ (in order to ensure the extensivity of the energy when the couplings have all the same sign).

Since the quantum Ising model is the basic short-ranged 
model in the field of zero-temperature quantum phase transitions \cite{sachdev},
its long-ranged version
\begin{eqnarray}
{\cal H}^{LR}_{QIpure}= - \sum_{n} h \sigma_n^z
 - \sum_{n<n'  }   \frac{J^x}{ \vert n-n' \vert^{1+\omega_x } }  \sigma_n^x \sigma_{n'}^x
\label{qipure}
\end{eqnarray}
has been also much studied in order to analyze how the critical properties 
depend upon the exponent $\omega_x$ \cite{dutta2001,werner2005,werner2005bis,norway2010,norway2012}.
The effects of random transverse-fields $h_n$  
\begin{eqnarray}
{\cal H}^{LR}_{QIrandom}= - \sum_{n} h_n \sigma_n^z
 - \sum_{n<n'  }   \frac{J^x}{ \vert n-n' \vert^{1+\omega_x } }  \sigma_n^x \sigma_{n'}^x
\label{hqisingrandom}
\end{eqnarray}
has also been studied recently \cite{strongLR1,strongLR2,strongLR3} (see also the related work \cite{epiLR} concerning
long-ranged epidemic models)
via the Strong Disorder Renormalization Group (SDRG) approach introduced by Daniel Fisher \cite{danielrtfic,danielantiferro,danielreview} (see the reviews \cite{strong_review1,strong_review2} for further references).


\subsection{ Dyson hierarchical version of long-ranged quantum spin chains }

\label{sec_dyson}

The hierarchical classical ferromagnetic Ising model introduced by Dyson \cite{dyson}
has been much studied by mathematicians
\cite{bleher,gallavotti,book,jona} and by physicists \cite{baker,mcguire,Kim,Kim77},
even from the dynamical point of view \cite{us_dysonferrodyn,c_dysondyn}.
As a consequence, the Dyson hierarchical versions of many other long ranged models
have been introduced :

(i) in the field of classical spin systems with quenched disorder, 
equilibrium properties have been analyzed for random fields Ising models \cite{randomfield,us_aval}
and for spin-glasses \cite{franz,castel_etal,castel_parisi,castel,angelini,us_rgsg1}, 
while the dynamical properties are discussed in \cite{us_rgsg2,c_dysondyn};

(ii) in the field of quantum localization, Dyson hierarchical models
have been considered for Anderson localization \cite{bovier,molchanov,krit,kuttruf,fyodorov,EBetOG,fyodorovbis,us_dysonloc}
and for many-body-localization \cite{mbl_emergent};

(iii) in the field of quantum spins,
the Dyson hierarchical version of the pure quantum Ising model of Eq. \ref{qipure}
has been studied via block-spin renormalization 
in order to analyze its critical properties \cite{c_dysontransverse}
and its entanglement properties for various bipartite partitions \cite{dyson_entang}.
The block-spin renormalization has been also applied
to the Dyson random transverse field Ising model \cite{c_dysontransverse}
and the quantum spin-glass in uniform transverse field \cite{c_dysonquantumSG}.

Even if Dyson hierarchical models have been introduced as theoretical toy models to simplify the renormalization analysis,
it is important to stress that they might become accessible in near-term cavity QED experiments \cite{dyson_experiment}.

\begin{figure}[h]
  \begin{center}
    \includegraphics[width=15cm]{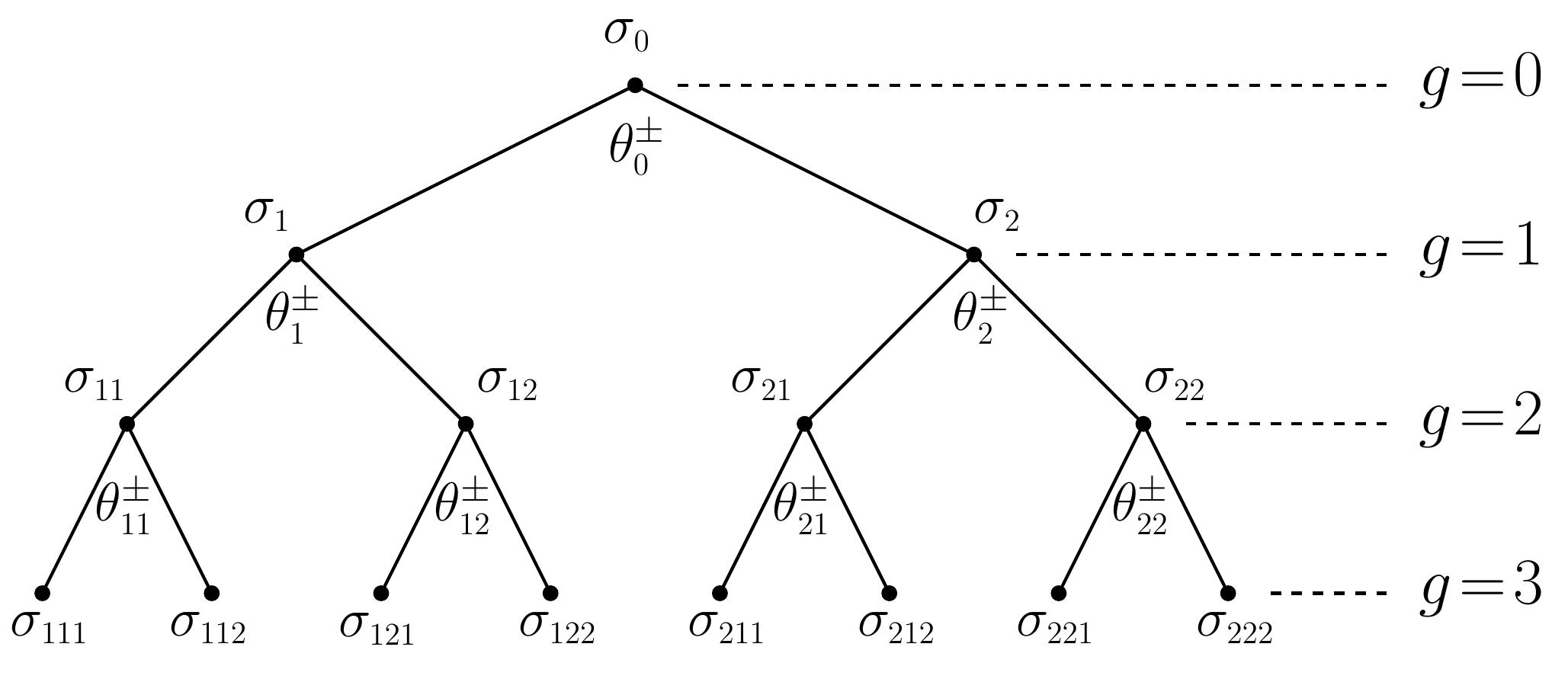}
  \end{center}
  \caption{ 
  The Dyson hierarchical models involving $2^3=8$ quantum spins $\sigma_{i_1 i_2 i_3}$ labelled by the three indices $i_1=1,2$ and $i_2=1,2$ and $i_3=1,2$ displayed on the last generation $g=3$ of the binary tree starting with the root at generation $g=0$ with the following interpretation : \\
  (a) The tree structure determines the amplitude of the couplings between the spins of the last generation $g=3$ (see Eq. \ref{dysong3pure}) \\
  (b) The spins of the other generations $g=2,1,0$ are introduced as renormalized spins during the block-renormalization procedure parametrized by the angles $\theta^{\pm}_{...}$ as described in section \ref{sec_treetensor} \\
  (c) For disordered models, the corresponding inhomogeneous Tree-Tensor-States with different angles $\theta^{\pm}_{...} $ at each node 
  are described in section \ref{sec_descending} \\
  (d) The optimization over the angles $\theta^{\pm}_{...} $ to obtain the lowest energy for disordered models is discussed in section \ref{sec_energy}. \\
  (e) For pure models, the symmetries yields that there is only one angle $\theta^+_{[g]}$ for each generation $g$ as discussed in section \ref{sec_pure}. 
  \\
  (e) For pure models at criticality, the additional scale invariance yields that there is only one angle $\theta^+$ for 
  the whole tree as discussed in section \ref{sec_criti}.
  }
  \label{Fig1}
\end{figure}

Let us now describe how Dyson hierarchical models are constructed in terms of the following binary tree structure 
(see Figure \ref{Fig1}) :
the generation $g=0$ contains the single site called the root;
the generation $g=1$ contains its two children labelled by the index $i_1=1,2$;
the generation $g=2$ contains the two children $i_2=1,2$ of each site $i_1=1,2$ of generation $g=1$,
and so on. So the generation $g$ contains $N_g=2^g$ sites labelled by the $g$ binary indices $(i_1,i_2,..,i_g)$
that indicate the whole line of ancestors up to the root at $g=0$.

The Dyson hierarchical version of the long-ranged Hamiltonian of Eqs \ref{hpt} and \ref{jpower}
is then defined for a chain of $N_G=2^G$ spins $\sigma_I$ 
labelled by the positions $I=(i_1,i_2,...,i_G)$ of the last generation $G$ of the tree structure 
by
\begin{eqnarray}
{\cal H}^{Dyson[G]}= - \sum_{I=(i_1,i_2,..,i_G)} h_I \sigma_I^{z}
- \sum_{I=(i_1,i_2,..,i_G) < I'=(i_1',i_2',..,i_G')} 
\sum_{a=x,y,z}  {\cal J}^a_{I,I'}         \sigma_I^{a} \sigma_{I'}^{a}
\label{dysonlca}
\end{eqnarray}
where the couplings $ {\cal J}^a_{I,I'}     $ have exactly the same power-law dependence as in Eq \ref{jpower}
\begin{eqnarray}
{\cal J}_{I,I'}^a = \frac{J^a_{I,I'}}{\left[ r(I,I')  \right]^{1+\omega_a } } 
\label{jpowertree}
\end{eqnarray}
but with respect to the ultrametric distance $r(I,I')$ on the tree defined 
in terms of the generation of their last common ancestor as follows (see Figure \ref{Fig1}).
Two sites $I=(I_c,i_{G-k}=1,F)$
and $I'=(I_c,i_{G-k}=2,F')$ that have in common the first $(G-k-1)$ indices 
$I_c=(i_1,..,i_{G-k-1 }) $, while they have different indices $i_{G-k}=1$ and $i_{G-k}=2$ at generation $(G-k)$,
are separated by the distance
\begin{eqnarray}
r(I=(I_c,i_{G-k}=1,F), I'=(I_c,i_{G-k}=2,F') ) \equiv 2^{k}
\label{rtree}
\end{eqnarray}
for any values of the remaining indices $F=(i_{G-k+1},..,i_G)$ and $F'=(i_{G-k+1}',..,i_G') $.
The minimal value $k=0$ corresponds to the distance $r=2^0=1$
between spins that have the same ancestor at position $I_c=(i_1,i_2,..,i_{G-1}) $ of the generation $(G-1)$
while they differ $i_G=1,2$ at generation $G$ (here $F$ and $F'$ are empty).
The maximal value $k=G-1$ corresponds to the distance $r=2^{G-1}=\frac{N_G}{2}$ 
between any spin belonging to the first half $i_1=1$ and any spin belonging to the second half $i_1=2$
(here $I_c$ is empty and their last common ancestor is the root $0$).

As an example, let us write the Dyson hierarchical Dyson version of the pure long-ranged quantum Ising chain of Eq. \ref{qipure}
for $G=3$ generations corresponding to $2^G=8$ spins $\sigma_{i_1 i_2 i_3}$ labelled by the three binary indices $i_1=1,2$
and $i_2=1,2$ and $i_3=1,2$ (see Figure \ref{Fig1})
\begin{eqnarray}
{\cal H}^{Dyson[G=3]}_{QIpure}=
&&  -  h \left( \sigma_{111}^{z} + \sigma_{112}^{z} 
 + \sigma_{121}^{z} + \sigma_{122}^{z} 
 + \sigma_{211}^{z} + \sigma_{212}^{z} 
 + \sigma_{221}^{z} + \sigma_{222}^{z} 
 \right)
\nonumber \\
&& - J^{x} \left( 
 \sigma_{111}^x \sigma_{112}^x
 +\sigma_{121}^x  \sigma_{122}^x 
 + \sigma_{211}^x  \sigma_{212}^x 
 + \sigma_{221}^x \sigma_{222}^x 
 \right)  
 \nonumber \\
&& -   \frac{ J^{x} }{  2^{(1+\omega_x)}  } 
 \left[ 
\left( \sigma_{111}^x + \sigma_{112}^x \right)  \left(\sigma_{121}^x +  \sigma_{122}^x \right)
+\left( \sigma_{211}^x + \sigma_{212}^x \right)  \left(\sigma_{221}^x +  \sigma_{222}^x \right)
 \right] 
 \nonumber \\
&& -  
 \frac{ J^{x} }{  2^{2(1+\omega_x)}  }
\left( \sigma_{111}^x + \sigma_{112}^x +\sigma_{121}^x +  \sigma_{122}^x \right)
\left( \sigma_{211}^x + \sigma_{212}^x +\sigma_{221}^x +  \sigma_{222}^x \right)  
\label{dysong3pure}
\end{eqnarray}

\section{  Elementary RG step for a block of two spins in terms of two angles}

\label{sec_treetensor}


\subsection{ Isometry $w_R$ for a block of two spins preserving the $(P,T)$ symmetries }

The basic block-spin RG step is implemented by the elementary coarse-graining isometry
$w_R$ between the two-dimensional Hilbert space $(\ket{\Uparrow}, \ket{\Downarrow})$ of the renormalized spin $\sigma_R^z$
and the two relevant states $\ket{ \psi_R^{\pm} }$
that are kept out of the four-dimensional Hilbert space $(\ket{\uparrow \uparrow}, \ket{\downarrow \downarrow },
\ket{\uparrow \downarrow}, \ket{\downarrow \uparrow })$
of its two children $\sigma_1^z$ and $\sigma_2^z$ 
\begin{eqnarray}
w_R
&& \equiv \ket{ \psi_R^+ }  \bra{\Uparrow } +   \ket{\psi_R^-  } \bra{\Downarrow }
\label{wisometry}
\end{eqnarray}
So the product
\begin{eqnarray}
w_R^{\dagger}w_R  && =\ket{\Uparrow }\bra{\Uparrow }
+ \ket{\Downarrow }  \bra{\Downarrow }=  \sigma_R^0
\label{wwdag}
\end{eqnarray}
is simply the identity operator $\sigma_R^0  $ of the Hilbert space of the renormalized spin, while the product
\begin{eqnarray}
w_R w_R^{\dagger} && = 
\ket{ \psi_R^+ } \bra{ \psi_R^+ }
 +  \ket{ \psi_R^- } \bra{ \psi_R^+ } 
\label{wdagw}
\end{eqnarray}
corresponds to the projector onto the subspace spanned by the two states $ \psi_R^{\pm} $ that are kept out of the four-dimensional Hilbert space of the two children.


In order to preserve the Parity,
the normalized ket $\ket{ \psi_R^+  }$
will be chosen as some linear combination of the two states of positive parity
$(\ket{\uparrow \uparrow}, \ket{\downarrow \downarrow }$
for the block of the two children.
Since the Time-Reversal symmetry imposes real coefficients,
the parametrization involves only a single angle $\theta_R^+$
\begin{eqnarray}
\ket{ \psi_R^+  }
&&
= \cos (\theta_R^+) \ 
\ket{\uparrow \uparrow }
 + \sin (\theta_R^+) \ 
\ket{\downarrow \downarrow  }
\label{psip}
\end{eqnarray}
Similarly, 
the ket $\ket{ \psi_R^-   }$
will be chosen as some real linear combination of the two states of negative parity 
$\ket{\uparrow \downarrow}, \ket{\downarrow \uparrow })$
for the block of the two children
 and involves only another single angle $\theta_R^-$
\begin{eqnarray}
\ket{ \psi_R^- }
&&
= \cos (\theta_R^-) \ 
\ket{\uparrow \downarrow }
 + \sin (\theta_R^-) \ 
\ket{\downarrow \uparrow }
\label{psim}
\end{eqnarray}

From the ascending block-spin renormalization  perspective,
Eqs \ref{psip} and \ref{psim} parametrize the representative states 
that are kept in each two-dimensional parity sector $P=\pm$ respectively.
From the descending perspective, Eqs \ref{psip} and \ref{psim}
can be interpreted as the Schmidt decompositions of the kept state $\ket{ \psi_R^{\pm}  } $,
where the two corresponding Schmidt singular values are given by
\begin{eqnarray}
\Lambda^{\pm}_1 && = \cos (\theta_R^{\pm})
\nonumber \\
\Lambda^{\pm}_2 && = \sin (\theta_R^{\pm})
\label{lambdatheta}
\end{eqnarray}
with the normalization of the corresponding weights
\begin{eqnarray}
 \left( \Lambda^{\pm}_1 \right)^2 +  \left( \Lambda^{\pm}_2 \right)^2 =\cos^2 (\theta_R^{\pm}) + \sin^2 (\theta_R^{\pm})=1
\label{lambdanorma}
\end{eqnarray}
So the two angles $\theta_R^{\pm} $ parametrize the entanglement properties of the two children in the kept states $\ket{ \psi_R^{\pm}  } $.


\subsection{ Ascending super-operator ${\cal A}_R$ and descending super-operator ${\cal D}_R  $ for the renormalization of operators }

The ascending super-operator ${\cal A}_R$ describes how the 
the 16 two-spin operators $ \sigma^{a_1=0,x,y,z}_{1} \sigma^{a_2=0,x,y,z}_{2}$ 
of the two children are projected onto the four Pauli operators 
 $\sigma^{a=0,x,y,z}_{R}  $ of the renormalized spin via the isometry $w_R$ of Eq. \ref{wisometry}
\begin{eqnarray}
 {\cal A}_R \left[\sigma^{a_1}_{1} \sigma^{a_2}_{2} \right]   \equiv 
w_R^{\dagger} \left( \sigma^{a_1}_{1} \sigma^{a_2}_{2} \right)  w_R   =
\sum_{a=0,x,y,z} F^{a}_{a_1,a_2}  \ \sigma^{a}_R
\label{ascendinglocal}
\end{eqnarray}
where the fusion coefficients
\begin{eqnarray}
 F^a_{a_1,a_2}  
=\frac{1}{2} {\rm Tr}_{\{ \sigma_R \}} \left(   \sigma^a_R   {\cal A} [ \sigma^{a_1}_{1} \sigma^{a_2}_{2} ]   \right)
=\frac{1}{2} {\rm Tr}_{\{ \sigma_R \}} \left(  \sigma^a_R w^{\dagger} \left( \sigma^{a_1}_{1} \sigma^{a_2}_{2} \right)  w \right)
\label{fusionFasc}
\end{eqnarray}
can be rewritten as
\begin{eqnarray}
 F^a_{a_1,a_2} 
 =\frac{1}{2} {\rm Tr}_{\{ \sigma_1,\sigma_2 \}} 
\left(  \sigma^{a_1}_{1} \sigma^{a_2}_{2}  w_R    \sigma^a_R w_R^{\dagger}  
 \right)
\label{coefK3}
\end{eqnarray}
As a consequence, the descending super-operator ${\cal D}_R$ 
that translates the four ancestor spin operators $\sigma^{a=0,x,y,z}_{R}  $ 
into operators for its two children involves the same fusion coefficients 
\begin{eqnarray}
 {\cal D}_R \left[\sigma^a_R \right]  
&& \equiv w_R  \sigma^a_R  w_R^{\dagger}
= \frac{1}{2} \sum_{a_1=0,x,y,z} \sum_{a_2=0,x,y,z} F^a_{a_1,a_2}
\ \sigma^{a_1}_{1} \sigma^{a_2}_{2}  
\label{descendinglocal}
\end{eqnarray}

Since the isometry $w_R$ preserves the Parity and the Time-Reversal symmetries,
the fusion rules respect the four symmetry sectors $(P=\pm 1,T=\pm 1)$ 
that allow to classify the operators 
according to their commutation $P=+1$ or anticommutation $P=-1$ with the Parity operator ${\cal P} $,
and according to their commutation $T=+1$ or anticommutation $T=-1$ with the  Time-Reversal operator ${\cal T}$.
For instance, since the sector $(P=+1,T=-1)$ is empty for operators concerning a single spin,
 the two operators of the two children belonging to the sector $(P=+1,T=-1)$ are projected out by the ascending super-operator ${\cal A}$
\begin{eqnarray}
 {\cal A}_R \left[\sigma^{x}_{1} \sigma^{y}_{2} \right]  && =  0
\nonumber \\
 {\cal A}_R \left[\sigma^{y}_{1} \sigma^{x}_{2} \right]  && =  0
\label{rgout}
\end{eqnarray}
and are never be produced by the ascending super-operator ${\cal D}_R$.

The fusion rules in the three other symmetry sectors are written
in the following subsections. In order to obtain simpler explicit expressions,
it will be convenient to replace the two angles $\theta_R^{\pm}$ by the two new angles
\begin{eqnarray}
\phi_R && \equiv   \frac{\pi}{2}- \theta^{+}_R - \theta^{-}_R 
\nonumber \\
 {\tilde \phi}_R && \equiv   - \theta_R^{+} + \theta_R^{-} 
\label{phi}
\end{eqnarray}
with the following simplified notations for their cosinus and sinus
\begin{eqnarray}
c_R && \equiv  \cos\left( \phi_R \right)
\nonumber \\
s_R && \equiv  \sin\left( \phi_R \right)
\nonumber \\
 {\tilde c}_R && \equiv  \cos \left(  {\tilde \phi}_R  \right)
\nonumber \\
{\tilde s}_R && \equiv  \sin \left(  {\tilde \phi}_R  \right)
\label{alphabeta}
\end{eqnarray}


\subsection{ Fusion rules for operators in the symmetry sector $(P=-1,T=+1)$ }

The action of the ascending super-operator ${\cal A}_R$
in the symmetry sector $(P=-1,T=+1)$ can only produce the 
operator $\sigma^{x}_R $ and the explicit computation yields the fusion coefficients
\begin{eqnarray}
 {\cal A}_R \left[\sigma^{x}_{1} \sigma^{0}_{2} \right]
    && 
= c_R   \ \sigma^{x}_R
\nonumber \\
 {\cal A}_R \left[\sigma^{0}_{1} \sigma^{x}_{2} \right]
  &&
=   {\tilde c}_R \ \sigma^{x}_R
\nonumber \\
 {\cal A}_R \left[\sigma^{x}_{1} \sigma^{z}_{2} \right]
   && =  {\tilde s}_R \  \sigma^{x}_R
\nonumber \\
 {\cal A}_R \left[\sigma^{z}_{1} \sigma^{x}_{2} \right]
&& 
= s_R  \  \sigma^{x}_R
\label{rgversx}
\end{eqnarray}
Reciprocally, these four operators 
 will appear in the application of the descending super-operator ${\cal D}_R$ to
 $\sigma^{x}_{R} $ with the same fusion coefficients according to Eq. \ref{descendinglocal}
\begin{eqnarray}
 {\cal D}_R \left[\sigma^x_R \right]  
=\frac{1}{2} 
\left[
 c_R \   \sigma^{x}_{1}  \sigma^{0}_{2}
+{\tilde c}_R \   \sigma^{0}_{1}  \sigma^{x}_{2}
+  {\tilde s}_R \  \sigma^{x}_{1}  \sigma^{z}_{2}
+ s_R \  \sigma^{z}_{1}  \sigma^{x}_{2}
\right]
\label{desfromx}
\end{eqnarray}
while the partial traces over a single child reduce to
\begin{eqnarray}
{\rm Tr}_{\{ \sigma_2 \} } \left(  {\cal D}_R \left[\sigma^{ x}_R \right]   \right)
&&  = c_R \ \sigma^{x}_{1} 
\nonumber \\
{\rm Tr}_{\{ \sigma_1 \} } \left(  {\cal D}_R \left[\sigma^{ x}_R \right]   \right)
&& = {\tilde c}_R \ \sigma^{x}_{2}
\label{descfromxtrace}
\end{eqnarray}
It is thus convenient to introduce the following notation
\begin{eqnarray}
\lambda^{x}_{R i} && \equiv 
 c_R \  \delta_{i,1}+  {\tilde c}_R \ \delta_{i,2} 
\label{lambdax}
\end{eqnarray}
to denote the scaling property of the single child operator $\sigma^{x}_{i}$
with respect to its ancestor operator $\sigma^{x}_R $.


\subsection{ Local fusion rules for operators in the symmetry sector $(P=-1,T=-1)$ }

Similarly, the action of the ascending super-operator ${\cal A}_R$
in the symmetry sector $(P=-1,T=-1)$ can only produce the 
operator $\sigma^{y}_R $ and the explicit computation yields the fusion coefficients
\begin{eqnarray}
 {\cal A}_R \left[\sigma^{y}_{1} \sigma^{0}_{2} \right]
&& =  {\tilde s}_R \ \sigma^{y}_R
\nonumber \\
 {\cal A}_R \left[\sigma^{0}_{1} \sigma^{y}_{2} \right]
&& = s_R  \  \sigma^{y}_R
\nonumber \\
 {\cal A}_R \left[\sigma^{y}_{1} \sigma^{z}_{2} \right]
    && 
= c_R \  \sigma^{y}_R
\nonumber \\
  {\cal A}_R \left[\sigma^{z}_{1} \sigma^{y}_{2} \right]
 && =  {\tilde c}_R \ \sigma^{y}_R
\label{rgversy}
\end{eqnarray}
Reciprocally, these four operators 
 will appear in the application of the descending super-operator to
 $\sigma^{y}_R $ with the same fusion coefficients according to Eq. \ref{descendinglocal}
\begin{eqnarray}
 {\cal D}_R \left[\sigma^{ y}_R \right]  
 =\frac{1}{2} 
\left[
  {\tilde s}_R \ \sigma^{y}_{1}  \sigma^{0}_{2}
+ s_R \  \sigma^{0}_{1}  \sigma^{y}_{2}
+ c_R \ \sigma^{y}_{1}  \sigma^{z}_{2}
+{\tilde c}_R \ \sigma^{z}_{1}  \sigma^{y}_{2}
\right]
\label{descfromy}
\end{eqnarray}
while the partial traces over a single child reduce to
\begin{eqnarray}
{\rm Tr}_{\{ \sigma_2 \} } \left(  {\cal D}_R \left[\sigma^{ y}_R \right]   \right)
&&  =  {\tilde s}_R \sigma^{y}_{1} 
\nonumber \\
{\rm Tr}_{\{ \sigma_1 \} } \left(  {\cal D}_R \left[\sigma^{ y}_R \right]   \right)
&& = s_R   \sigma^{y}_{2}
\label{descfromytrace}
\end{eqnarray}
Again it is convenient to introduce the following notation
\begin{eqnarray}
\lambda^{y}_{Ri} && \equiv 
  {\tilde s}_R \  \delta_{i,1}+ s_R \ \delta_{i,2} 
\label{lambday}
\end{eqnarray}
to denote the scaling property of the single child operator $\sigma^{y}_{i}$
with respect to its ancestor operator $\sigma^{y}_R $.


\subsection{ Local fusion rules for operators in the symmetry sector $(P=+1,T=+1)$ }

In the symmetry sector $(P=+1,T=+1)$,
the action of the ascending super-operator ${\cal A}_R $
can only involve 
the two operators $\sigma^{a=0,z}_R $.
The identity $\sigma^{0}_{1} \sigma^{0}_{2} $  of the children space
is projected onto the identity $\sigma^{0}_R$ of the ancestor space
as a consequence of Eq \ref{wwdag}
\begin{eqnarray}
 {\cal A}_R \left[\sigma^{0}_{1} \sigma^{0}_{2} \right]
= \sigma^{0}_R
\label{azerozero}
\end{eqnarray}
while the parity $\sigma^{z}_{1} \sigma^{z}_{2} $ of the block of the two children
is projected onto the parity $ \sigma^{z}_R$ of the ancestor
\begin{eqnarray}
 {\cal A}_R \left[\sigma^{z}_{1} \sigma^{z}_{2} \right]
= \sigma^{z}_R
\label{azz}
\end{eqnarray}
The remaining operators are projected onto the following linear combinations of 
the two operators $\sigma^{0}_R $ and $\sigma^{z}_R $ 
\begin{eqnarray}
 {\cal A}_R \left[\sigma^{z}_{1} \sigma^{0}_{2} \right]
  && = 
 s_R   {\tilde c}_R \ \sigma^{0}_R
+ c_R  {\tilde s}_R \ \sigma^{z}_R
\nonumber \\
 {\cal A} \left[\sigma^{0}_{1} \sigma^{z}_{2} \right]  && = 
 c_R  {\tilde s}_R \ \sigma^{0}_R
+ s_R   {\tilde c}_R \  \sigma^{z}_R
\nonumber \\
 {\cal A}_R \left[\sigma^{x}_{1} \sigma^{x}_{2} \right]
 &&
 = c_R   {\tilde c}_R \ \sigma^{0}_R
-  s_R  {\tilde s}_R \  \sigma^{z}_R
\nonumber \\
  {\cal A}_R \left[\sigma^{y}_{1} \sigma^{y}_{2} \right]
&&
 = 
  s_R {\tilde s}_R \ \sigma^{0}_R
- c_R   {\tilde c}_R \ \sigma^{z}_R
\label{afour}
\end{eqnarray}


Reciprocally, Eq. \ref{descendinglocal} yields
that the application of the descending super-operator $ {\cal D}_R$
to
$\sigma^{z}_R $ reads
(only the block identity $\sigma^{0}_{1}  \sigma^{0}_{2} $ does not appear)
\begin{eqnarray}
 {\cal D}_R \left[\sigma^{ z}_R \right]  
 =\frac{1}{2} 
\left[  c_R {\tilde s}_R
\  \sigma^{z}_{1}  \sigma^{0}_{2}
+ s_R {\tilde c}_R
\  \sigma^{0}_{1}  \sigma^{z}_{2}
\right]
 +\frac{1}{2} 
\left[\sigma^{z}_{1}  \sigma^{z}_{2}
- s_R {\tilde s}_R
\  \sigma^{x}_{1}  \sigma^{x}_{2}
- c_R  {\tilde c}_R
\  \sigma^{y}_{1}  \sigma^{y}_{2}
\right]
\label{descz}
\end{eqnarray}
while the partial traces over a single child reduce to
\begin{eqnarray}
{\rm Tr}_{\{ \sigma_2 \} } \left( {\cal D}_R \left[\sigma^{ z}_R \right]    \right)
&&  = 
c_R {\tilde s}_R \ \sigma^{z}_{1}  
\nonumber \\
{\rm Tr}_{\{ \sigma_1 \} } \left( {\cal D}_R \left[\sigma^{ z}_R \right]    \right)
&& =  
 s_R {\tilde c}_R \  \sigma^{z}_{2}
\label{descfromztrace}
\end{eqnarray}
Again it is convenient to introduce the following notation
\begin{eqnarray}
\lambda^{z}_{Ri} && \equiv 
  c_R   {\tilde s}_R \   \delta_{i,1}+s_R  {\tilde c}_R \   \delta_{i,2} 
\label{lambdaz}
\end{eqnarray}
to denote the scaling property of the single child operator $\sigma^{z}_{i}$
with respect to its ancestor operator $\sigma^{z}_R $.


Finally, Eq. \ref{descendinglocal} yields
that the application of the descending super-operator $ {\cal D}_R$
to the identity $\sigma^{0}_R$ of the ancestor space
reads
(only the block parity $\sigma^{z}_{1}  \sigma^{z}_{2} $ does not appear)
\begin{eqnarray}
  {\cal D}_R \left[\sigma^{ 0}_R \right]  
 =\frac{1}{2} 
\left[ \sigma^{0}_{1}  \sigma^{0}_{2}
+ s_R {\tilde c}_R
\  \sigma^{z}_{1}  \sigma^{0}_{2}
+ c_R  {\tilde s}_R
\  \sigma^{0}_{1}  \sigma^{z}_{2}\right]
 +\frac{1}{2} 
\left[ c_R  {\tilde c}_R
\  \sigma^{x}_{1}  \sigma^{x}_{2}
+ s_R {\tilde s}_R 
\  \sigma^{y}_{1}  \sigma^{y}_{2}
\right]
\label{desc0} 
\end{eqnarray}
while the partial traces over a single child reduce to
\begin{eqnarray}
{\rm Tr}_{\{ \sigma_2 \} } \left(  {\cal D}_R \left[\sigma^{ 0}_R \right]    \right)
&&  = 
\sigma^{0}_{1}  
+ s_R {\tilde c}_R \  \sigma^{z}_{1}  
\nonumber \\
{\rm Tr}_{\{ \sigma_1 \} } \left(  {\cal D}_R \left[\sigma^{ 0}_R \right]    \right)
&& =   \sigma^{0}_{2}
+ c_R  {\tilde s}_R \   \sigma^{z}_{2}
\label{descfrom0trace}
\end{eqnarray}
Although the meaning is different from the three scaling factors $\lambda^{a=x,y,z}_{Ri}$
introduced above, it will be nevertheless convenient to introduce
\begin{eqnarray}
\lambda^{0}_{Ri} && \equiv 
s_R  {\tilde c}_R \    \delta_{i,1}+ c_R   {\tilde s}_R \delta_{i,2} 
\label{lambda0}
\end{eqnarray}
to denote the scaling property of the single child operator $\sigma^{z}_{i}$
with respect to its ancestor identity $\sigma^{0}_R $.


\section{  Tree-Tensor-State with its one and two spins reduced density matrices }

\label{sec_descending}

In this section, the goal is to construct the simplest inhomogeneous 
Tree-Tensor-States for disorder spin chains with Parity and Time-Reversal symmetries,
while the case of homogeneous Tree-Tensor-States for pure spin chains 
is postponed to the sections \ref{sec_pure} and \ref{sec_criti}
where their supplementary symmetries will be taken into account.

\subsection{ Assembling the elementary isometries to build the whole Tree-Tensor-State of parity $P=+$  }

The traditional block-spin renormalization procedure based on blocks of two spins 
can be summarized as follows in terms of the tree notations introduced in the subsection \ref{sec_dyson}.
The initial chain of $N_G=2^G$ spins $\sigma^{[G]}_{i_1,i_2,..,i_G}$ belonging to the last generation $G$
will be first renormalized into a chain of $N_{G-1}=2^{G-1}=\frac{N_G}{2}$ spins $\sigma^{[G-1]}_{i_1,i_2,..,i_{G-1}}$ of generation $(G-1)$.
This procedure will be then iterated up to the last RG step where there will be a single spin $\sigma^{[g=0]}$
at the root corresponding to generation $g=0$.

The correspondence between the ket $\ket{ \Psi^{[g]} }  $ for the chain of generation $g$ containing $N_g=2^g$ spins
and the ket $\ket{ \Psi^{[g+1]} }  $ for the chain of generation $(g+1)$ containing $N_{g+1}=2^{g+1}$ spins
\begin{eqnarray}
\ket{ \Psi^{[g+1]} } = W^{[g]}\ket{ \Psi^{[g]} } 
\label{wpket}
\end{eqnarray}
is described by the global isometry $W^{[g]} $ made of the tensor product 
over the $2^g$ positions $I=(i_1,i_2,..,i_g )$
of the elementary isometries $w_I $ of Eq. \ref{wisometry} parametrized by two angles $\theta^{\pm}_I $
\begin{eqnarray}
W^{[g]}= \prod_{ I=(i_1,i_2,..,i_g ) }  w_I 
\label{wgtot}
\end{eqnarray}
At generation $g=0$, the state of the single spin $\sigma_0^z$ at the root of the tree 
represents the Parity $P$ of the whole chain.
We will focus on the positive parity sector $P=+$ corresponding to the initial ket
\begin{eqnarray}
\ket{ \Psi^{[0] } } =  \ket{\sigma^{z}_0=P=+ }  
\label{psi0}
\end{eqnarray}
The iteration of the rule of Eq. \ref{wpket} will then generate
a Tree-Tensor-State of parity $P=+$ for the chain of generation $g$ containing $N_g=2^g$ spins
\begin{eqnarray}
\ket{ \Psi^{[g]} }  = W^{[g-1]}  \ket{ \Psi^{[g-1] } } =W^{[g-1]}  W^{[g-2]} ...  W^{[1]} W^{[0]}\ket{ \Psi^{[0]} }
\label{iterket}
\end{eqnarray}
Since the elementary isometry $w_{I'} $ at generation $g'$ and position $I'$
is parametrized by the two angles $\theta^{[g',I']\pm} $, 
the global isometry $W^{[g']} $ for the $2^{g'}$ sites $I'$ of the generation $g'$ involves $2 \times 2^{g'} $ angles,
except for the generation $g'=0$ where only the angle $\theta^{+}_0  $ will appear for the initial condition of Eq. \ref{psi0}.
So the total number of parameters involved in the Tree-Tensor-State $\ket{ \Psi^{[g]} }  $
of generation $g$ containing $N_g=2^g$ spins grows only linearly with respect to $N_g$
\begin{eqnarray}
{\cal N}^{Parameters}_g =  1+2 \times \sum_{g'=1}^{g-1} 2^{g'} = 2 (2^g-1) -1 = 2 N_g-3
\label{npara}
\end{eqnarray}

As explained around Eq \ref{lambdatheta}, these angles $\theta^{\pm}_I  $ parametrize the
hierarchical entanglement 
at different levels.
The consequences of this tree-tensor structure for the entanglement of various bipartite partitions
have been studied in detail in \cite{dyson_entang}  on the specific case of the pure Dyson quantum Ising model.
In the following subsections, we focus instead on the consequences
for the one-point and two-point reduced density matrices
that allow to compute any one-spin and two-spins observable.


\subsection{ Recursion for the full density matrices via the full descending super-operator ${\cal D}^{[g]}$ }

The full density matrix for the chain at generation $g$
\begin{eqnarray}
\rho^{[g]}  \equiv \ket{ \Psi^{[g]} } \bra{ \Psi^{[g]} }  
\label{rhochain}
\end{eqnarray}
satisfies the recurrence involving the global descending super-operator ${\cal D}^{[g]} $
\begin{eqnarray}
\rho^{[g+1]}  
&& = W^{[g]} \rho^{[g]}   (W^{[g]})^{\dagger} \equiv {\cal D}^{[g]} \left[ \rho^{[g]}  \right]
\label{recrhog}
\end{eqnarray}
while the initial condition at generation $g=0$ corresponds to the projector (Eq \ref{psi0})
\begin{eqnarray}
\rho^{[0]}  \equiv \ket{ \Psi^{[0] } } \bra{ \Psi^{[0] } }    = \frac{\sigma_0^0+   \sigma^{z}_0}{2}
\label{rho0}
\end{eqnarray}
Since the density matrix $\rho^{[g]} $ of generation $g$ 
can be expanded into the Pauli basis of the $2^g$ spins $ \sigma^{a_I=0,x,y,z}_{I}$,
one just needs to know how to
apply the descending super-operator ${\cal D}^{[g]} $ to products of Pauli matrices
\begin{eqnarray}
{\cal D}^{[g]} \left[\prod_{I=(i_1,I_2,..,i_g)} \sigma^{a_I}_{I} \right]  
 = 
 \prod_{I=(i_1,I_2,..,i_g)} {\cal D}_I \left[\sigma^{a_I}_{I} 
\right]
\label{descendingexpli}
\end{eqnarray}
where the properties of the elementary descending super-operator ${\cal D}_I $ of Eq. \ref{descendinglocal}
have been described in detail in the previous section.


\subsection{ Parametrization of 1 and 2 spins reduced density matrices in terms of magnetizations and correlations }

In order to compute all the one-spin and two-spins observables, 
one just needs the one-spin and two-spins reduced density matrices.
Since the initial condition of Eq. \ref{rho0} belongs to the symmetry sector $(P=+,T=+)$,
the full density matrices of Eq. \ref{recrhog} obtained by the successive application of 
the global descending super-operator ${\cal D}^{[g]} $ are also in the sector $(P=+,T=+)$,
and the partial traces over some spins will also preserve this symmetry sector.
As a consequence, the single-spin reduced density matrix $\rho_{I}$ of generation $g$ at position $I$ can be parametrized as
\begin{eqnarray}
\rho_{I}  =  \frac{ \sigma^{0}_I+m_{I} \sigma^{z}_I}{2}
\label{rho1spin}
\end{eqnarray}
where the coefficient $1/2$ of the identity $\sigma^{0}_{I}  $ is fixed by the normalization 
\begin{eqnarray}
{\rm Tr}_{\{ I\} }(\rho_{I} )  = 1
\label{rhognorma}
\end{eqnarray}
while $m_{I} $ represents the magnetization at site $I$ of generation $g$
\begin{eqnarray}
m_{I} ={\rm Tr}_{\{ I \} } (\sigma^{z}_{I}\rho_{I} )
\label{rhogmagn}
\end{eqnarray}
Similarly, the reduced density matrix $\rho_{I,I'}$ of two spins at positions
$I$ and $I'$ of generation $g$
can only involve the six two-spin operators  
 of the sector $(P=+,T=+)$
 and can be thus parametrized as
\begin{eqnarray}
\rho_{I,I'} &&  = 
\frac{\sigma^{0}_{I}  \sigma^{0}_{I'}}{4} 
+ m_{I}
 \frac{ \sigma^{z}_{I}  \sigma^{0}_{I'}}{4} 
+m_{I'}
 \frac{ \sigma^{0}_{I}  \sigma^{z}_{I'}}{4} 
+C^{z}_{I,I'}
 \frac{\sigma^{z}_{I}  \sigma^{z}_{I'}}{4} 
+C^{x}_{I,I'}
\frac{ \sigma^{x}_{I}  \sigma^{x}_{I'}}{4} 
+C^{y}_{I,I'}
 \frac{\sigma^{y}_{I}  \sigma^{y}_{I'}}{4} 
\label{rhogparameters}
\end{eqnarray}
where the three first coefficients are fixed by the compatibility
 with the one-point reduced density matrices of Eq. \ref{rho1spin},
while the three last coefficients $C^{a=x,y,z}_{I,I'}$ represent the two-points $xx$ , $yy$ and $zz$ correlations
\begin{eqnarray}
C^{a}_{I,I'} ={\rm Tr}_{\{ I,I' \} } (\sigma^{a}_{I}\sigma^{a}_{I'}\rho_{I,I'} )
\label{rhogcorre}
\end{eqnarray}


\subsection{ Recursions for the one-point magnetizations and the two-point correlations  }

The application of the local descending super-operator $ {\cal D}_I $ 
to the reduced density matrix $\rho_{I}$
of the single site $I$ of generation $g$ of Eq. \ref{rho1spin}
produces the following reduced density matrix of its two children $(I1,I2)$ of generation $(g+1)$
using Eqs \ref{descz} and \ref{desc0}
\begin{eqnarray}
 \rho_{I1,I2}   = {\cal D}_I [\rho_{I} ] 
&& = \frac{1}{2} 
\left[  {\cal D}_I[ \sigma^{0}_{I} ] 
+ m_{I}
  {\cal D}_I[ \sigma^{z}_{I} ]
\right]
\label{rho2sameblock} \\
&&  = 
 \frac{\sigma^{0}_{I1}  \sigma^{0}_{I2}}{4} 
+ \left( s_I {\tilde c}_I
+ m_{I} c_I {\tilde s}_I
\right)
 \frac{\sigma^{z}_{I1}  \sigma^{0}_{I2}}{4} 
+\left( c_I  {\tilde s}_I 
+ m_{I} s_I {\tilde c}_I\right)
 \frac{ \sigma^{0}_{I1}  \sigma^{z}_{I2}}{4} 
 \nonumber \\ &&
+  m_{I}
 \frac{\sigma^{z}_{I1}  \sigma^{z}_{I2}}{4} 
+\left(  c_I  {\tilde c}_I
- m_{I}  s_I {\tilde s}_I
\right)
 \frac{ \sigma^{x}_{I1}  \sigma^{x}_{I2}}{4} 
+ \left(  s_I {\tilde s}_I 
- m_{I} c_I  {\tilde c}_I
\right)
 \frac{\sigma^{y}_{I1}  \sigma^{y}_{I2}}{4} 
\nonumber 
\end{eqnarray}
The identification with the parametrization of Eq. \ref{rhogparameters}
 yields the following affine recursions for the magnetizations 
where one recognizes the coefficients $\lambda^{a=0,z}_{Ii} $ with the index $i=1,2$ for the two children
introduced in Eqs \ref{lambdaz} and \ref{lambda0}
\begin{eqnarray}
m_{I 1} && = s_I {\tilde c}_I
+ m_{I} c_I {\tilde s}_I=\lambda^{0}_{I1}+\lambda^{z}_{I1} m_{I} 
\nonumber \\
m_{I 2} && = c_I  {\tilde s}_I 
+ m_{I} s_I {\tilde c}_I=\lambda^{0}_{I2}+\lambda^{z}_{I2} m_{I} 
\label{recmu}
\end{eqnarray}
and gives how the correlations between the two children of the same ancestor are produced
in terms of the coefficients $\lambda^{a=x,y}_{Ii} $ introduced in Eqs \ref{lambdax} and \ref{lambday}
\begin{eqnarray}
C^{x}_{I1,I2} && =  c_I  {\tilde c}_I
- m_{I}  s_I {\tilde s}_I = \lambda^{x}_{I1}  \lambda^{x}_{I2} - \lambda^{y}_{I1}  \lambda^{y}_{I2}  m_{I}  
\nonumber \\
C^{y}_{I1,I2} && = s_I {\tilde s}_I 
- m_{I} c_I  {\tilde c}_I = \lambda^{y}_{I1}  \lambda^{y}_{I2} - \lambda^{x}_{I1}  \lambda^{]x}_{I2}  m_{I} 
\nonumber \\
C^{z}_{I1,I2} && = m_{I}
\label{recorresameblock}
\end{eqnarray}

The application of the descending super-operator $ {\cal D}^{[g]} $ to the reduced density matrix $\rho_{I,I'}$
of two different sites $I \ne I'$ of generation $g$ of Eq. \ref{rhogparameters}
will produce the four-sites reduced density matrix for their children $(I1,I2)$ and $(I'1,I'2)$ of generation $(g+1)$
\begin{eqnarray}
 \rho_{I1,I2,I'1,I'2}   = {\cal D}^{[g]} [\rho_{I,I'} ] 
&&  = \frac{1}{4} 
\left[  {\cal D}_I[ \sigma^{0}_{I} ] 
{\cal D}_{I'}[ \sigma^{0}_{I'}]
+ m_{I}
  {\cal D}_I[ \sigma^{z}_{I} ] 
{\cal D}_{I'}[ \sigma^{0}_{I'}]
+m_{I'}
  {\cal D}_I[ \sigma^{0}_{I} ]
 {\cal D}_{I'}[ \sigma^{z}_{I'}]
\right]
\nonumber \\
&&  + \frac{1}{4} 
\left[
C^{z}_{I,I'}
 {\cal D}_I[ \sigma^{z}_{I} ] 
{\cal D}_{I'}[ \sigma^{z}_{I'}]
+
C^{x}_{I,I'}
  {\cal D}_I[ \sigma^{x}_{I} ] 
{\cal D}_{I'}[ \sigma^{x}_{I'}]
+C^{y}_{I,I'}
 {\cal D}_I[ \sigma^{y}_{I} ]
 {\cal D}_{I'}[ \sigma^{y}_{I'}]
\right]
\label{rhofour}
\end{eqnarray}
One then needs to take the trace over one child in each block to obtain the reduced density matrices
of the two remaining children
\begin{eqnarray}
 \rho_{I1,I'1}   && ={\rm Tr}_{\{ I2,I'2  \} } \left(  \rho_{I1,I2,I'1,I'2}  \right) 
\nonumber \\
\rho_{I1,I'2}   && ={\rm Tr}_{\{ I2,I'1  \} } \left(  \rho_{I1,I2,I'1,I'2}  \right) 
\nonumber \\
\rho_{I2,I'1}   && ={\rm Tr}_{\{ I1,I'2  \} } \left(   \rho_{I1,I2,I'1,I'2}  \right) 
\nonumber \\
\rho_{I2,I'2}   && ={\rm Tr}_{\{ I1,I'1  \} } \left(   \rho_{I1,I2,I'1,I'2} \right) 
\label{rho2oddodd}
\end{eqnarray}
Using the partial traces over a single child in each block computed before in Eqs 
\ref{descfromxtrace} \ref{descfromytrace}  \ref{descfromztrace} \ref{descfrom0trace},
one obtains the following rules for the two-point correlations
between the children of different blocks.
The $xx$ and $yy$ correlations are governed by the following multiplicative factorized rules as long as $I \ne I'$
\begin{eqnarray}
C^{x}_{Ii,I'i'} && =\lambda^{x}_{Ii}    \lambda^{x}_{I'i'}    C^{x}_{I,I'} 
\label{recorrexx}
\end{eqnarray}
\begin{eqnarray}
C^{y}_{Ii,I'i'} && =\lambda^{y}_{Ii}    \lambda^{y}_{I'i'}    C^{y}_{I,I'}
\label{recorreyy}
\end{eqnarray}
The recursions for the $zz$ correlations involve four terms
\begin{eqnarray}
C^{z}_{Ii,I'i'} && =
\lambda^{0}_{Ii}    \lambda^{0}_{I'i'} 
+\lambda^{z}_{Ii}    \lambda^{0}_{I'i'}    m_{I} 
+ \lambda^{0}_{Ii}    \lambda^{z}_{I'i'} m_{I'} 
+\lambda^{z}_{Ii}    \lambda^{z}_{I'i'}     C^{z}_{I,I'} 
\label{recorrezz}
\end{eqnarray}
so that it is more convenient to consider the simpler multiplicative factorized rule
satisfied by the connected correlations for $I \ne I'$ using Eq. \ref{recmu}
\begin{eqnarray}
\left( C^{z}_{Ii,I'i'} - m_{Ii}  m_{I'i'} \right) 
&& =
\lambda^{z}_{Ii}    \lambda^{z}_{I'i'}   
\left( C^{z}_{I,I'} - m_{I}  m_{I'} \right)
\label{recorrezzconnected}
\end{eqnarray}


\subsection{ Explicit solution for the one-point magnetizations  }

The initial condition for the magnetization at generation $g=0$ is given by the chosen parity $P=+$ (Eq \ref{rho0})
\begin{eqnarray}
m_0 && =+1
\label{recmuP0}
\end{eqnarray}
The first iterations of the affine recursion of Eq. \ref{recmu} 
give for the generations $g=1$ and $g=2$
\begin{eqnarray}
m_{i_1} && =\lambda^{0}_{i_1}+\lambda^{z}_{i_1}  
\nonumber \\
m_{i_1,i_2} && = \lambda^{0}_{i_1,i_2}+   \left( \lambda^{0}_{i_1}+\lambda^{z}_{i_1}   \right)\lambda^{z}_{i_1,i_2}
\label{recmuP12}
\end{eqnarray}
More generally, one obtains that 
the magnetization at position $(i_1,...,i_g)$ of generation $g$ reads
\begin{eqnarray}
m_{i_1,...,i_g} 
 = \lambda^{0}_{i_1,...i_g}
+ \sum_{g''=1}^{g-1} \lambda^{0}_{i_1,...i_{g''}}
\left( \prod_{g'=g''+1}^{g}  \lambda^{z}_{i_1,...i_{g'}}  \right)
 + 
\left( \prod_{g'=1}^{g}  \lambda^{z}_{i_1,...i_{g'}}  \right) 
\label{solumu}
\end{eqnarray}
The first term involving the single scaling factor $\lambda^{0}_{i_1,...i_g}$
is already enough to produce a finite magnetization, while the last term
involving the $g$ scaling factors up to the initial condition of the root 
will be exponentially small.


\subsection{ Explicit solution for the two-points correlations  }

The $xx$ correlations between two sites $(I,1,i_{G-k+1},..,i_G)$
and $(I,2,i_{G-k+1}'..,i_G')$ that have their last common ancestor at the
position $I=(i_1,..,i_{G-k-1 }) $ of generation $g=G-k-1$
and that are thus at distance $r_k=2^k$ (Eq \ref{rtree}) on the tree
satisfy the recursion of Eq \ref{recorrexx} 
as long as they are apart
\begin{eqnarray}
C^{x}_{(I,1,i_{G-k+1},..,i_G),(I,2,i_{G-k+1}',..,i_G') } 
&& =    C^{x}_{ I1,I2} 
\left( \prod_{g'=G-k+1 }^{G }  \lambda^{x}_{I,1,i_{G-k+1},..,i_{g'}}  
  \lambda^{x}_{I,2,i_{G-k+1}',..,i_{g'}'}  
\right)
\label{recorrexxmu}
\end{eqnarray}
while the remaining correlation $C^{x}_{ I1,I2} $ at generation $(G-k)$ is given by Eq. \ref{recorresameblock}
in terms of the magnetization $ m_{I}  $ of their common ancestor
\begin{eqnarray}
C^{x}_{I1,I2} && = \lambda^{x}_{I1}  \lambda^{x}_{I2} 
- \lambda^{y}_{I1}  \lambda^{y}_{I2}  m_{I}  
\label{cxfirstancestor}
\end{eqnarray}
and will thus be finite.
As a consequence, the decay of the correlation of Eq. \ref{recorrexxmu} with respect to the distance $r=2^k$
will be governed by the two strings of the $k$ scaling factors $\lambda^{x}_{....}$ 
leading to their last common ancestor.

Similarly, the $yy$ correlations are given by
\begin{eqnarray}
C^{y}_{(I,1,i_{G-k+1},..,i_G),(I,2,i_{G-k+1}',..,i_G') } 
&& =  C^{y}_{ I1,I2} 
\left( \prod_{g'=G-k+1 }^{G }  \lambda^{y}_{I,1,i_{G-k+1},..,i_{g'}}  
  \lambda^{y}_{I,2,i_{G-k+1}',..,i_{g'}'}  
\right)
\label{recorreyymu}
\end{eqnarray}
with
\begin{eqnarray}
C^{y}_{I1,I2} && = \lambda^{y}_{I1}  \lambda^{y}_{I2} 
- \lambda^{x}_{I1}  \lambda^{x}_{I2}  m_{I}  
\label{cyfirstancestor}
\end{eqnarray}
Again, the decay of the correlation of Eq. \ref{recorreyymu} with respect to the distance $r=2^k$
will be governed by the two strings of the $k$ scaling factors $\lambda^{y}_{....}$ 
leading to their last common ancestor $I$.

Finally, the $zz$ connected correlations satisfying Eq \ref{recorrezzconnected} read
\begin{eqnarray}
&& \left( C^{z}_{(I,1,i_{G-k+1},..,i_G),(I,2,i_{G-k+1}',..,i_G') } 
-  m_{(I,1,i_{G-k+1},..,i_G)} m_{(I,2,i_{G-k+1}',..,i_G)'} 
\right)
 \nonumber \\ &&
 = \left(  C^{z}_{ (I1),(I2)} - m_{I1}   m_{I2}  \right)
\left( \prod_{g'=G-k +1}^{G }  \lambda^{z}_{I,1,i_{G-k+1},..,i_{g'}}  
  \lambda^{z}_{I,2,i_{G-k+1}',..,i_{g'}}  
\right)
\label{recorrezzmu}
\end{eqnarray}
where the remaining connected correlation at generation $(G-k)$ is given by Eqs 
\ref{recorresameblock} and \ref{recmu}
\begin{eqnarray}
C^{z}_{I1,I2}  - m_{I1} m_{I2}= 
m_{I}
- \left(\lambda^{0}_{I1}+\lambda^{z}_{I1} m_{I} \right)
\left( \lambda^{0}_{I2}+\lambda^{z}_{I2} m_{I} \right)
\nonumber \\
\label{czfirstancestor}
\end{eqnarray}
in terms of the magnetization $ m_{I}  $ of their common ancestor.


\section{ Energy of the Tree-Tensor-State and optimization of its parameters }

\label{sec_energy}

Up to now, we have only used the Parity and the Time-Reversal symmetries
to build the simplest inhomogeneous Tree-Tensor-States and analyze its general properties.
In the present section, we take into account the specific form of the Hamiltonian,
in order to evaluate the energy of the Tree-Tensor-State and to optimize its parameters.

\subsection{ Energy of the Tree-Tensor-State in terms of the magnetizations and the correlations at generation $G$}

For the Dyson Hamiltonian ${\cal H}^{Dyson[G]} $ of Eq. \ref{dysonlca} that contains only one-body and two-body terms,
the energy ${\cal E}^{[G]} $ of the Tree-Tensor-State $\ket{ \Psi^{[G]} } $ 
\begin{eqnarray}
{\cal E}^{[G]} \equiv \bra{\Psi^{[G]}} {\cal H}^{Dyson[G]} \ket{ \Psi^{[G]} }  = {\rm Tr}_{\{G\}} ({\cal H}^{[G]} \rho^{[G]} ) 
\label{energie}
\end{eqnarray}
involves only the one-body and the two-body reduced density matrices of Eqs \ref{rho1spin}
and \ref{rhogparameters}. It can be thus rewritten in terms of the magnetizations $m_{I} $
and of the two-point correlations $C^{a=x,y,z}_{I,I'} $ as
\begin{eqnarray}
{\cal E}^{[G]}= - \sum_{I=(i_1,i_2,..,i_G)} h_I m_I
- \sum_{I=(i_1,i_2,..,i_G) < I'=(i_1',i_2',..,i_G')} 
\sum_{a=x,y,z}  \frac{J^a_{I,I'}}{\left[ r(I,I')  \right]^{1+\omega_a } }     C^{a}_{I,I'}  
\label{edysonlca}
\end{eqnarray}
All the Tree-Tensor-State parameters are contained in the magnetizations $m_{I} $
and of the two-point correlations $C^{a=x,y,z}_{I,I'} $ computed in the previous section,
and one could thus try to write the optimization equations by deriving Eq. \ref{edysonlca}
with respect to the various parameters. However, in order to isolate more clearly the role of each parameter,
it is more convenient 
to consider how the energy of the Tree-Tensor-State of
Eq. \ref{edysonlca} can be rewritten in terms of the variables associated to any other generation $g$.

\subsection{ Energy of the Tree-Tensor-State in terms of the properties at generation $g$ }

Using the recurrence for the density matrices of Eq. \ref{rhochain},
the energy of Eq. \ref{energie} can be rewritten via 
the relation between ascending and descending super-operators
as 
the energy for the spin chain at generation $(G-1)$
\begin{eqnarray}
{\cal E}^{[G]} =  {\rm Tr}_{\{G\}} \left({\cal H}^{[G]} {\cal D}^{[G-1]} \left[ \rho^{[G-1]}  \right] \right) 
= {\rm Tr}_{\{G-1\}} ({\cal H}^{[G-1]}  \rho^{[G-1]}  ) \equiv {\cal E}^{[G-1]} 
\label{energiefirst}
\end{eqnarray}
of the renormalized Hamiltonian
\begin{eqnarray}
{\cal H}^{[G-1]}  = {\cal A}^{[G-1]} \left[ {\cal H}^{[G]} \right] =  ( W^{[G-1]})^{\dagger}   {\cal H}^{[G]}   W^{[G-1]}
\label{hrgfirst}
\end{eqnarray}
obtained via the application of the full ascending super-operator ${\cal A}^{[G-1]} $ to the initial Hamiltonian ${\cal H}^{[G]} $.
More generally, it is convenient to introduce the renormalized Hamiltonian ${\cal H}^{[g]} $
at any generation $g$ via the recurrence
\begin{eqnarray}
{\cal H}^{[g]} && = {\cal A}^{[g]} [ {\cal H}^{[g+1]} ] 
\label{hptpa}
\end{eqnarray}
Since the ascending super-operators preserve the Parity and Time-Reversal symmetries,
the renormalized Hamiltonian ${\cal H}^{[g]} $ can be parametrized in terms 
renormalized fields $h_I  $ and in terms renormalized couplings $ J^{a}_{I,I'} $,
as well as a constant term $E^{[g]} $
\begin{eqnarray}
{\cal H}^{[g]}=E^{[g]} - \sum_{I=(i_1,i_2,..,i_g)} h_I \sigma_I^{z}
- \sum_{I=(i_1,i_2,..,i_g) < I'=(i_1',i_2',..,i_g')} \sum_{a=x,y,z}   \frac{J^{a}_{I,I'}}{\left[ r(I,I')  \right]^{1+\omega_a } } 
   \sigma_I^{a}
 \sigma_{I'}^{a}
\label{dysonlcag}
\end{eqnarray}
The energy ${\cal E}^{[G]} $ of Eq. \ref{energie} that was computed for the last generation $G$
can be then rewritten as
the energy ${\cal E}^{[g]} $ of the renormalized Hamiltonian ${\cal H}^{[g]} $ at any generation $g$
in terms of the magnetizations $m_{I} $
and correlations $C^{a=x,y,z}_{I,I'} $ of the generation $g$
\begin{eqnarray}
{\cal E}^{[G]}  ={\cal E}^{[g]}  \equiv  {\rm Tr}_{\{g\}} ({\cal H}^{[g]}  \rho^{[g]}  ) 
 =E^{[g]} - \sum_{I=(i_1,i_2,..,i_g)} h_I m_I
- \sum_{I=(i_1,i_2,..,i_g) < I'=(i_1',i_2',..,i_g')} \sum_{a=x,y,z}  \frac{J^{a}_{I,I'}}{\left[ r(I,I')  \right]^{1+\omega_a } }   C^{a}_{I,I'}  
\label{energieg}
\end{eqnarray}


\subsection{ Renormalization rules for the parameters of the Hamiltonian }

The renormalization rules for the parameters of the renormalized Hamiltonian of Eq. \ref{dysonlcag}
can be derived via the application of the ascending super-operator in Eq. \ref{hptpa}.
An alternative derivation uses the identification 
between the energy computed at generation $g$ with Eq. \ref{energieg}
and the energy computed at the next generation $(g+1)$ 
\begin{eqnarray}
{\cal E}^{[g]} ={\cal E}^{[g+1]}  && = E^{[g+1]}- \sum_{I=(i_1,i_2,..,i_g)} 
\sum_{i } h_{Ii} m_{Ii} 
- \sum_{I=(i_1,i_2,..,i_g)} 
\sum_{a=x,y,z}  J^{a}_{I1,I2}   C^{a}_{I1,I2}  
\nonumber \\
&& - \sum_{I=(i_1,i_2,..,i_g) < I'=(i_1',i_2',..,i_g')} 
\sum_{i } \sum_{i' }
\sum_{a=x,y,z} \frac{J^{a}_{Ii,I'i'}}{\left[ 2 r(I,I')  \right]^{1+\omega_a } }  
   C^{a}_{Ii,I'i'}    
\label{energiegp1}
\end{eqnarray}
Plugging the recursions for the magnetization (Eq \ref{recmu})
and for the two-point correlations in the same block  (Eq \ref{recorresameblock})
or in two different blocks (Eqs \ref{recorrexx}, \ref{recorreyymu} and \ref{recorrezz})
into Eq. \ref{energiegp1}
\begin{eqnarray}
&& {\cal E}^{[g+1]}   = E^{[g+1]} - \sum_{I=(i_1,i_2,..,i_g)} 
\left[ \sum_{i } h_{Ii} (\lambda^{0}_{Ii}+\lambda^{z}_{Ii} m_{I} )
+ \sum_{I=(i_1,..,i_g)}   J^{z}_{I1,I2}   m_{I} \right]
\label{energiegnext}
\\
&&- \sum_{I=(i_1,..,i_g)} 
\left[   J^{x}_{I1,I2}
  ( \lambda^{x}_{I1}  \lambda^{x}_{I2} - \lambda^{y}_{I1}  \lambda^{y}_{I2}  m_{I}  )
+   J^{y}_{I1,I2} 
 (\lambda^{y}_{I1}  \lambda^{y}_{I2} - \lambda^{x}_{I1}  \lambda^{x}_{I2}  m_{I} )
\right]
\nonumber \\
&& - \sum_{I=(i_1,..,i_g) < I'=(i_1',i_2',..,i_g')} 
\sum_{i } \sum_{i' }
\left[ \frac{J^{x}_{Ii,I'i'}}{\left[ 2 r(I,I')  \right]^{1+\omega_x } }  
\lambda^{x}_{Ii}    \lambda^{x}_{I'i'}    C^{x}_{I,I'} 
+
\frac{J^{y}_{Ii,I'i'}}{\left[ 2 r(I,I')  \right]^{1+\omega_y } }  
\lambda^{y}_{Ii}    \lambda^{y}_{I'i'}    C^{y}_{I,I'}  \right]
\nonumber \\
&& - \sum_{I=(i_1,..,i_g) < I'=(i_1',..,i_g')} 
\sum_{i } \sum_{i' }
\frac{J^{z}_{Ii,I'i'}}{\left[ 2 r(I,I')  \right]^{1+\omega_z } }  
(\lambda^{0}_{Ii}    \lambda^{0}_{I'i'} 
+ \lambda^{z}_{Ii} \lambda^{0}_{I'i'}   m_{I} 
+ \lambda^{0}_{Ii}    \lambda^{z}_{I'i'} m_{I'} 
+\lambda^{z}_{Ii}    \lambda^{z}_{I'i'}    C^{z}_{I,I'} ) 
\nonumber 
\end{eqnarray}
one obtains via the identification with Eq. \ref{energieg}
the following renormalization rules.

The renormalized couplings $J_{I,I'}^{a}$
are simply given by linear combinations of the four corresponding couplings between their children
$(I1,I2)$ and $(I'1,I'2)$ of generation $(g+1)$
\begin{eqnarray}
J^{a}_{I,I'}
   = 
\sum_{i=1,2}\sum_{i'=1,2}
\frac{ \lambda^{a}_{Ii}  
\lambda^{a}_{I'i'}    }{2^{1+\omega_a } }  
J^{a}_{Ii,I'i'}
\label{rgj}
\end{eqnarray}
As a consequence, if some coupling component $a=x,y,z$ is not present in the initial Hamiltonian, it will not be generated via renormalization.

The renormalized field $h_I $ involves local terms coming from the two fields 
$h_{I1}$ and $h_{I2} $ of its children and
the three couplings between its two children $J_{I1,I2}^{a=x,y,z}   $, 
but also long-ranged terms coming from all the $z$-couplings between one child $(I1)$ 
or the other $(I2)$ with other children from other blocks $I' \ne I$ :
\begin{eqnarray}
   h_I 
 &&  =  h_{I1} \lambda^{z}_{I1} 
+  h_{I2} \lambda^{z}_{I2}
+   J^{z}_{I1,I2} 
-   J^{x}_{I1,I2} 
     \lambda^{y}_{I1}  \lambda^{y}_{I2}  
-   J^{y}_{I1,I2} 
  \lambda^{x}_{I1}  \lambda^{x}_{I2} 
\nonumber \\
&& + \sum_{I'=(i_1',i_2',..,i_g')>I} 
\sum_{i } \sum_{i' }
\frac{J^{z}_{Ii,I'i'}}{\left[ 2 r(I,I')  \right]^{1+\omega_z } }  
 \lambda^{z}_{Ii} \lambda^{0}_{I'i'}  
 + \sum_{I'=(i_1',i_2',..,i_g')<I } 
\sum_{i } 
 \sum_{i' } 
\frac{J^{z}_{I'i',Ii}}{\left[ 2 r(I,I')  \right]^{1+\omega_z } }  
 \lambda^{0}_{I'i'}    \lambda^{z}_{Ii} 
\label{rgh}
\end{eqnarray}
As a consequence, even if the fields are not present in the initial Hamiltonian, 
they will be generated via renormalization. 
In addition, the presence of $z$-couplings $J^{z}_{..} $ in the initial Hamiltonian leads
to qualitatively different RG rules with non-local contributions,
while if the $z$ couplings $J^{z}_{..} =0 $ are not present in the initial Hamiltonian, the RG rule for the fields $h_I$
are local and only involves the fields and couplings of its two children.

Finally, the renormalization of the constant contribution 
involves the random fields, the $x$-couplings and the $y$-couplings between the two spins $(I1,I2)$ 
of the blocks, as well as the $z$-couplings between spins $(I1,I2)$ and $(I'1,I'2)$
belonging to different blocks $I<I'$
\begin{eqnarray}
  E^{[g]} &&  = E^{[g+1]} 
 - \sum_{I=(i_1,i_2,..,i_g)} 
\left( 
 h_{I1} \lambda^{0}_{I1}
+  h_{I2} \lambda^{0}_{I2}
+   J^{x}_{I1,I2} 
   \lambda^{x}_{I1}  \lambda^{x}_{I2}
+   J^{y}_{I1,I2} 
 \lambda^{y}_{I1}  \lambda^{y}_{I2} \right)
\nonumber \\
&& - \sum_{I=(i_1,i_2,..,i_g) < I'=(i_1',i_2',..,i_g')} 
\sum_{i=1,2 } \sum_{i' =1,2}
\frac{J^{z}_{Ii,I'i'}}{\left[ 2 r(I,I')  \right]^{1+\omega_z } }  
\lambda^{0}_{Ii}    \lambda^{0}_{I'i'} 
\label{rgconstant} 
\end{eqnarray}
So again the presence of $z$-couplings $J^{z}_{..} $ in the initial Hamiltonian leads
to qualitatively different RG rules with non-local contributions.


\subsection{ How the energy of the Tree-Tensor-State depends on the parameters of generation $g$ }

We have seen above how the 
 energy of the Tree-Tensor-State can be computed at any generation $g=G,G-1,..,0$
\begin{eqnarray}
 {\cal E}^{[G]}  && = {\cal E}^{[g]} 
\label{energylastlast}
\end{eqnarray}
At generation $g=G$, the energy of Eq. \ref{edysonlca}
involves the fields and couplings of the initial Hamiltonian $ {\cal H}^{[G]}  $,
while the whole dependence on the parameters of the Tree-Tensor-State
is contained in the magnetization $m_I$ and correlations $C^{a}_{I,I'}   $ for the $N_G=2^G$ spins of generation $G$.
At generation $g=0$ where there is a single spin left $\sigma_0$ with magnetization unity $m_0=1$,
the energy involves instead the constant term $E^{[g=0]} $ and the renormalized field $h_0  $
obtained at the end of the renormalization procedure for the Hamiltonian
\begin{eqnarray}
 {\cal E}^{[G]} =  {\cal E}^{[0]}  =  E^{[0]} - h_0 
\label{energyzero}
\end{eqnarray}
So here the whole dependence on the parameters of the Tree-Tensor-State
is contained in the renormalized parameters $(E^{[0]} , h_0 )$ of the Hamiltonian at generation $g=0$.

At any intermediate generation $g=1,..,G-1$, the dependence on the parameters of the Tree-Tensor-State
of the energy ${\cal E}^{[g]} $ of Eq. \ref{energieg} is divided in two parts :
 the magnetizations $m_I$ and the correlations $C^{a}_{I,I'}   $ of generation $g$ 
only involve the Tree-Tensor-State parameters of smaller generations $g'=0,..,g-1$,
while the renormalized parameters $(E^{[g]},h_{I}, {\cal J}^{a}_{I,I'} )$ of the Hamiltonian 
of generation $g$ only involve the Tree-Tensor-State parameters of bigger generations $g'=g,..,G-1$.

As a consequence, the dependence of the energy with respect to the 
parameters of generation $g$ can be seen in Eq. \ref{energiegnext} via the scaling factors 
$\lambda^{I]a=0,x,y,z}_{Ii} $ of generation $g$, 
while the dependence with respect to smaller generations $g'=0,..,g-1$
is contained in the magnetizations $m_I$ and the correlations $C^{a}_{I,I'}   $
of generation $g$, and the dependence with respect to bigger generations $g'=g+1,..,G-1 $
is contained in the renormalized parameters of the Hamiltonian of generation $(g+1)$.

Since the general case with the the $z$-couplings leads to somewhat heavy expressions
for the disordered models described by inhomogeneous Tree-Tensor-States,
it is more instructive to focus now on the simpler models without $z$-couplings,
while we will return to the general case with the three type of couplings $a=x,y,z$
in the next sections concerning pure models.


\subsection{ Optimization of the parameters  of the Tree-Tensor-State for the case without $z$-couplings }

For the case without $z$-couplings, the discussion of the previous subsection yields
that the optimization of Eq. \ref{energiegnext}  
with respect to the angle $\phi_I $ of Eq \ref{phi} at position $I=(i_1,..,i_g)$ of generation $g$ reads
\begin{eqnarray}
0= \frac{\partial  {\cal E}^{[g+1]}  }{ \partial \phi_I  } = 
&&
-   h_{I1} 
\left(  \frac{ \partial \lambda^{0}_{I1}}{ \partial \phi_I  } 
+\frac{\partial \lambda^{z}_{I1}  }{ \partial \phi_I  } m_{I} 
\right)
-   h_{I2} 
\left(  \frac{ \partial \lambda^{0}_{I2}}{ \partial \phi_I  } 
+\frac{\partial \lambda^{z}_{I2}  }{ \partial \phi_I  } m_{I} 
\right)
\nonumber \\
&&-  J^{x}_{I1,I2}
  \left( 
\frac{\partial (\lambda^{x}_{I1}  \lambda^{x}_{I2})  }{ \partial \phi_I  }
 - \frac{\partial (\lambda^{y}_{I1}  \lambda^{y}_{I2})  }{ \partial \phi_I  }
  m_{I}  
\right)
-   J^{y}_{I1,I2} 
 \left(\frac{ \partial (\lambda^{y}_{I1}  \lambda^{y}_{I2} ) }{ \partial \phi_I  } 
- \frac{\partial (\lambda^{x}_{I1}  \lambda^{x}_{I2}  ) }{ \partial \phi_I  }
 m_{I} \right)
\nonumber \\
&&  - \sum_{i }
\frac{\partial \lambda^{x}_{Ii}   }{ \partial \phi_I  } 
\left[  \sum_{I'>I} 
 \sum_{i' }
 \frac{J^{x}_{Ii,I'i'}    C^{x}_{I,I'} }{\left[ 2 r(I,I')  \right]^{1+\omega_x } }  
   \lambda^{x}_{I'i'}  
+
 \sum_{I'<I} 
 \sum_{i' }
 \frac{J^{x}_{I'i',Ii}   C^{x}_{I',I} }{\left[ 2 r(I,I')  \right]^{1+\omega_x } }  
  \lambda^{x}_{I'i'}   
\right]
\nonumber \\
&& 
- \sum_{i }
\frac{\partial \lambda^{y}_{Ii}   }{ \partial \phi_I  } 
\left[  \sum_{I'>I} 
 \sum_{i' }
 \frac{J^{y}_{Ii,I'i'}    C^{y}_{I,I'} }{\left[ 2 r(I,I')  \right]^{1+\omega_y } }  
   \lambda^{y}_{I'i'}  
+
 \sum_{I'<I} 
 \sum_{i' }
 \frac{J^{y}_{I'i',Ii}   C^{y}_{I',I} }{\left[ 2 r(I,I')  \right]^{1+\omega_y } }  
  \lambda^{y}_{I'i'}   
\right]
\label{energiegnextderi}
\end{eqnarray}
and the analogous equation with respect to the other angle $ {\tilde \phi}_I$.

Using the explicit forms of the scaling factors 
$\lambda^{a=0,x,y,z}_{Ii} $ for $i=1,2$ of Eqs \ref{lambdax} \ref{lambday} \ref{lambdaz} \ref{lambda0},
Eq. \ref{energiegnextderi} becomes
\begin{eqnarray}
0= \frac{\partial  {\cal E}^{[g+1]}  }{ \partial \phi_I  } && = 
-   h_{I1} \left(  c_I  {\tilde c}_I- s_I   {\tilde s}_I  m_{I} \right)
-   h_{I2} \left(  - s_I   {\tilde s}_I+  c_I   {\tilde c}_I  m_{I} \right)
\nonumber \\
&&+  J^{x}_{I1,I2}  \left( s_I    {\tilde c}_I  + c_I {\tilde s}_I   m_{I}  \right)
-    J^{y}_{I1,I2}  \left({\tilde s}_I  c_I+ s_I   {\tilde c}_I m_{I} \right)
\nonumber \\
&& +s_I 
\left[  \sum_{I'>I} 
 \sum_{i' }
 \frac{J^{x}_{I1,I'i'}    C^{x}_{I,I'} }{\left[ 2 r(I,I')  \right]^{1+\omega_x } }  
   \lambda^{x}_{I'i'}  
+
 \sum_{I'<I} 
 \sum_{i' }
 \frac{J^{x}_{I'i',I1}   C^{x}_{I',I} }{\left[ 2 r(I,I')  \right]^{1+\omega_x } }  
  \lambda^{x}_{I'i'}   
\right]
\nonumber \\
&& 
- c_I 
\left[  \sum_{I'>I} 
 \sum_{i' }
 \frac{J^{y}_{I2,I'i'}    C^{y}_{I,I'} }{\left[ 2 r(I,I')  \right]^{1+\omega_y } }  
   \lambda^{y}_{I'i'}  
+
 \sum_{I'<I} 
 \sum_{i' }
 \frac{J^{y}_{I'i',I2}   C^{y}_{I',I} }{\left[ 2 r(I,I')  \right]^{1+\omega_y } }  
  \lambda^{y}_{I'i'}   
\right]
\label{energiegnextderi1}
\end{eqnarray}
Similarly, the optimization equation with respect to the angle ${\tilde \phi}_I$ yields
\begin{eqnarray}
0= \frac{\partial  {\cal E}^{[g+1]}  }{ \partial {\tilde \phi}_I  } && = 
-   h_{I1} \left(  - s_I  {\tilde s}_I+c_I   {\tilde c}_I m_{I} \right)
-   h_{I2} \left(  c_I   {\tilde c }_I -  s_I {\tilde s}_I m_{I} \right)
\nonumber \\
&&+  J^{x}_{I1,I2}  \left(  c_I    {\tilde s}_I +    s_I {\tilde c}_I   m_{I}  \right)
-    J^{y}_{I1,I2}  \left( {\tilde c}_I  s_I+ c_I   {\tilde s}_I m_{I} \right)
\nonumber \\
&& + {\tilde s}^{[g,I]}
\left[  \sum_{I'>I} 
 \sum_{i' }
 \frac{J^{x}_{I2,I'i'}    C^{x}_{I,I'} }{\left[ 2 r(I,I')  \right]^{1+\omega_x } }  
   \lambda^{x}_{I'i'}  
+
 \sum_{I'<I} 
 \sum_{i' }
 \frac{J^{x}_{I'i',I2}   C^{x}_{I',I} }{\left[ 2 r(I,I')  \right]^{1+\omega_x } }  
  \lambda^{x}_{I'i'}   
\right]
\nonumber \\
&& 
-  {\tilde c}_I  
\left[  \sum_{I'>I} 
 \sum_{i' }
 \frac{J^{y}_{I1,I'i'}    C^{y}_{I,I'} }{\left[ 2 r(I,I')  \right]^{1+\omega_y } }  
   \lambda^{y}_{I'i'}  
+
 \sum_{I'<I} 
 \sum_{i' }
 \frac{J^{y}_{I'i',I1}   C^{y}_{I',I} }{\left[ 2 r(I,I')  \right]^{1+\omega_y } }  
  \lambda^{y}_{I'i'}   
\right]
\label{energiegnextderi2}
\end{eqnarray}

It is simpler to replace these two equations
by their sum
\begin{eqnarray}
0 && = (1+m_{I}  ) 
\left[ -   (h_{I1} + h_{I2} )  (c_I  {\tilde c}_I- s_I  {\tilde s}_I)
+ \left(  J^{x}_{I1,I2} -  J^{y}_{I1,I2}  \right)  (s_I    {\tilde c}_I + c_I    {\tilde s}_I) \right]
\nonumber \\
&& +(s_I +{\tilde s}_I)
\left[  \sum_{I'>I} 
 \sum_{i' }
 \frac{J^{x}_{I1,I'i'}    C^{x}_{I,I'} }{\left[ 2 r(I,I')  \right]^{1+\omega_x } }  
   \lambda^{x}_{I'i'}  
+
 \sum_{I'<I} 
 \sum_{i' }
 \frac{J^{x}_{I'i',I1}   C^{x}_{I',I} }{\left[ 2 r(I,I')  \right]^{1+\omega_x } }  
  \lambda^{x}_{I'i'}   
\right]
\nonumber \\
&& 
- (c_I +  {\tilde c}_I  )
\left[  \sum_{I'>I} 
 \sum_{i' }
 \frac{J^{y}_{I2,I'i'}    C^{y}_{I,I'} }{\left[ 2 r(I,I')  \right]^{1+\omega_y } }  
   \lambda^{y}_{I'i'}  
+
 \sum_{I'<I} 
 \sum_{i' }
 \frac{J^{y}_{I'i',I2}   C^{y}_{I',I} }{\left[ 2 r(I,I')  \right]^{1+\omega_y } }  
  \lambda^{y}_{I'i'}   
\right]
\label{energiegnextderisomme}
\end{eqnarray}
and by their difference
\begin{eqnarray}
0= && (1-m_{I}  ) 
\left[ -   ( h_{I1} - h_{I2}  )  \left(  c_I  {\tilde c}_I + s_I  {\tilde s}_I \right)
+  \left( J^{x}_{I1,I2} + J^{y}_{I1,I2}  \right)\left( s_I    {\tilde c}_I - c_I    {\tilde s}_I \right) \right]
\nonumber \\
&& +(s_I - {\tilde s}_I)
\left[  \sum_{I'>I} 
 \sum_{i' }
 \frac{J^{x}_{I1,I'i'}    C^{x}_{I,I'} }{\left[ 2 r(I,I')  \right]^{1+\omega_x } }  
   \lambda^{x}_{I'i'}  
+
 \sum_{I'<I} 
 \sum_{i' }
 \frac{J^{x}_{I'i',I1}   C^{x}_{I',I} }{\left[ 2 r(I,I')  \right]^{1+\omega_x } }  
  \lambda^{x}_{I'i'}   
\right]
\nonumber \\
&& 
- (c_I -  {\tilde c}_I  )
\left[  \sum_{I'>I} 
 \sum_{i' }
 \frac{J^{y}_{I2,I'i'}    C^{y}_{I,I'} }{\left[ 2 r(I,I')  \right]^{1+\omega_y } }  
   \lambda^{y}_{I'i'}  
+
 \sum_{I'<I} 
 \sum_{i' }
 \frac{J^{y}_{I'i',I2}   C^{y}_{I',I} }{\left[ 2 r(I,I')  \right]^{1+\omega_y } }  
  \lambda^{y}_{I'i'}   
\right]
\label{energiegnextdiff}
\end{eqnarray}

\subsection{ Comparison with the usual block-spin RG rules based on the diagonalization 
of the intra-Hamiltonian  }

The usual block-spin RG rules based on the diagonalization 
of the intra-Hamiltonian in each block (see Appendix \ref{app_diago})
are recovered if one neglects the correlations $C^{[g]a=x,y}_{I,I'} \to 0$
that correspond to the future RG steps in the formula given above.
Then Eq \ref{energiegnextderisomme} simplifies into
\begin{eqnarray}
0= && 
-   (h_{I1} + h_{I2} )  (c_I  {\tilde c}_I- s_I  {\tilde s}_I)
+ \left(  J^{x}_{I1,I2} -  J^{y}_{I1,I2}  \right)  (s_I    {\tilde c}_I + c_I    {\tilde s}_I)
\nonumber \\
&& = -   (h_{I1} + h_{I2} )  \cos( \phi_I +{\tilde \phi}_I  ) 
+ \left(  J^{x}_{I1,I2} -  J^{y}_{I1,I2}  \right)  \sin( \phi_I +{\tilde \phi}_I  ) 
\label{sommesimpli}
\end{eqnarray}
for the sum of the two angles (Eq \ref{phi})
\begin{eqnarray}
\phi_I +  {\tilde \phi}_I   =   \frac{\pi}{2}- 2\theta^{+}_I
\label{phisum}
\end{eqnarray}
in agreement with Eq. \ref{thetas} of the Appendix.
Similarly, Eq. \ref{energiegnextdiff} simplifies into
\begin{eqnarray}
0= && 
-   ( h_{I1} - h_{I2}  )  \left(  c_I  {\tilde c}_I + s_I  {\tilde s}_I \right)
+  \left( J^{x}_{I1,I2} + J^{y}_{I1,I2}  \right)\left( s_I    {\tilde c}_I - c_I    {\tilde s}_I \right)
\nonumber \\
&& 
= -   ( h_{I1}- h_{I2}  ) \cos( \phi_I -{\tilde \phi}_I  ) 
+  \left( J^{x}_{I1,I2} + J^{y}_{I1,I2}  \right) \sin( \phi_I -{\tilde \phi}_I  ) 
\label{diffsimpli}
\end{eqnarray}
for the difference of the two angles (Eq \ref{phi})
\begin{eqnarray}
\phi_I - {\tilde \phi}_I  =   \frac{\pi}{2}- 2\theta^-_I 
\label{phidiff}
\end{eqnarray}
in agreement with Eq. \ref{thetaa} of the Appendix.

In summary, with respect to the block-spin RG rules of Eqs \ref{sommesimpli} and \ref{diffsimpli}
based on the diagonalization of the renormalized intra-Hamiltonian in each block
that contains the isometries of bigger generations $g'=g+1,..,G-1$,
the variational optimization of the whole Tree-Tensor-State of Eqs \ref{energiegnextderisomme} and \ref{energiegnextdiff}
requires to take into account the magnetizations $m_I$ and the 
correlations $C^{a=x,y}_{I,I'} $ that contain the isometries of smaller generations $g'=0,..,g-1$.


\section{ Homogeneous Tree-Tensor-States for the pure Dyson hierarchical models }

\label{sec_pure}

Up to now, we have considered inhomogeneous Tree-Tensor-States for disordered Dyson hierarchical quantum spin models.
In this section, we turn to the case of pure Dyson hierarchical quantum spin models, where 
their supplementary symmetries need to be taken into account in the Tree-Tensor description.

\subsection{ Supplementary symmetries of the pure Dyson models }

When the fields $h_I$ and the couplings $ J^a_{I,I'}$
of the Dyson Hamiltonian (Eqs \ref{dysonlca} \ref{jpowertree})
are uniform 
\begin{eqnarray}
h_I=h
\nonumber \\
J^a_{I,I'}=J^a
\label{hpur}
\end{eqnarray}
one needs to take into account two supplementary symmetries for the 
choice of the isometries of Eq. \ref{wisometry}.
The first symmetry concerns the equivalence between the various branches of the tree,
so that the two angles $\theta^{\pm}_I$ of Eqs \ref{psip} and \ref{psim}
 will only depend on the generation $g$
but not on the position $I=(i_1,..,i_g)$ anymore
\begin{eqnarray}
\theta^{\pm}_I =\theta^{\pm}_{[g]}
\label{thetapur}
\end{eqnarray}
The second symmetry concerns the equivalence of the two children of a given ancestor.
In the parity sector $P=+$, the ket of Eq. \ref{psip} is symmetric with respect to the two children
for any value of the angle $\theta^{+}_{[g]} $. However in the parity sector $P=-$, the ket of Eq. \ref{psim}
  is symmetric with respect to the two children only for the value
\begin{eqnarray}
\theta^{-}_{[g]}=\frac{\pi}{4}
\label{symenfants}
\end{eqnarray}

As a consequence, the number of parameters of Eq. \ref{npara} needed for the inhomogeneous Tree-Tensor-States
corresponding to disordered models
is now reduced to the number $G$ of generations for the homogeneous Tree-Tensor-States
corresponding to pure models
\begin{eqnarray}
{\cal N}^{Parameters}_{pure} =   \sum_{g=0}^{G-1} 1 = G
\label{nparapure}
\end{eqnarray}

The two symmetries of Eq. \ref{thetapur} and \ref{symenfants} 
yields that the two angles $\phi_{[g]} $ and ${\tilde \phi}_{[g]} $ of Eq. \ref{phi} now coincide  
\begin{eqnarray}
\phi_{[g]}  =  {\tilde \phi}_{[g]}  =  \frac{\pi}{4}- \theta^{+}{\tilde \phi}_{[g]}
\label{phipure}
\end{eqnarray}
so the
coefficients of Eq. \ref{alphabeta} reduce to
\begin{eqnarray}
c_{[g]} && \equiv  \cos\left( \phi_{[g]}   \right) \equiv {\tilde c}_{[g]}
\nonumber \\
s_{[g]} && \equiv  \sin\left( \phi_{[g]}  \right) \equiv {\tilde s}_{[g]}
\label{alphabetapure}
\end{eqnarray}
and the scaling factors of Eqs \ref{lambdax} \ref{lambday} \ref{lambdaz} \ref{lambda0} simplify into
\begin{eqnarray}
\lambda^{x}_{[g]} && \equiv  c_{[g]} 
\nonumber \\
\lambda^{y}_{[g]} && \equiv  s_{[g]} 
\nonumber \\
\lambda^{z}_{[g]} && \equiv   c_{[g]}   s_{[g]} 
\nonumber \\
\lambda^{0}_{[g]} && \equiv   c_{[g]}   s_{[g]} 
\label{muxyz0}
\end{eqnarray}


\subsection{ Explicit solution for the one-point magnetizations  }

The magnetization now only depends on the generation $g$.
The recursion of Eq. \ref{recmu} simplifies into
\begin{eqnarray}
m_{[g+1]} && =  c_{[g]}   s_{[g]}  (1+ m_{[g]})
\label{recmupure}
\end{eqnarray}
and the solution of Eq. \ref{solumu} 
reduces to
\begin{eqnarray}
m_{[g]}
&& =  \sum_{g''=0}^{g-1} \left( \prod_{g'=g''}^{g-1}  c_{[g']}   s_{[g']}   \right) 
 + 
\left( \prod_{g'=0}^{g-1}   c_{[g']}   s_{[g']}   \right) 
\label{solumupure}
\end{eqnarray}

\subsection{ Explicit solutions for the two-point correlations  }

The two-point correlations between two sites of generation $G$
now only depend on the generation $g=G-k-1$ of their last common ancestor
i.e. on their corresponding distance $r_k=2^k$ on the tree.
Eqs \ref{recorresameblock} give the values of the correlations at distance $r_0=2^0=1$
as a function of the magnetization given in Eq. \ref{solumupure}
\begin{eqnarray}
C^{x}_{[g+1]} (r_0=1)  && = c_{[g]}^2 -  s_{[g]}^2   m_{[g]}  
\nonumber \\
C^{y}_{[g+1]} (r_0=1) && =  s_{[g]}^2 -  c_{[g]}^2   m_{[g]}  
\nonumber \\
C^{z}_{[g+1]} (r_0=1) && = m_{[g]}
\label{Cpure1}
\end{eqnarray}
while Eqs \ref{recorrexx}, \ref{recorreyy} \ref{recorrezzconnected} correspond to the following recursions for $k \geq 0$
\begin{eqnarray}
C^{x}_{[g+1]} (2r_k=2^{k+1}) && =  c_{[g]}^2  C^{x}_{[g]}(r_k=2^k)
\nonumber \\
C^{y}_{[g+1]} (2r_k=2^{k+1}) && = s_{[g]}^2    C^{y}_{[g]}(r_k=2^k)
\nonumber \\
\left( C^{z}_{[g+1]} (2r_k=2^{k+1}) -m_{[g+1]}^2 \right) 
&& =
 c_{[g]}^2 s_{[g]}^2
\left(  C^{z}_{[g]}(r_k=2^k) - m_{[g]}^2 \right)
\label{recorrepure}
\end{eqnarray}

The solutions of Eqs \ref{recorrexxmu} \ref{recorreyymu} and \ref{recorrezzmu}  reduce to
\begin{eqnarray}
C^{x}_{[G]} (r_k=2^k) 
&& =
\left( \prod_{g'=G-k }^{G-1 }   c_{[g']}    \right)
 \left[   c_{[G-k-1]}^2  -  s_{[G-k-1]}^2    m_{[G-k-1]}  \right]
\nonumber \\
C^{y}_{[G]} (r_k=2^k) 
&& =
\left( \prod_{g'=G-k }^{G-1 } s_{[g']}^2 \right)
\left[ 
  s_{[G-k-1]}^2  -  c_{[G-k-1]}^2    m_{[G-k-1]}  
\right]
\nonumber \\
 C^{z}_{[G]} (r_k=2^k) 
-   m_{[G]}^2
&& =
\left( \prod_{g'=G-k }^{G-1 }   c_{[g']}^2  s_{[g']}^2 
\right)
 \left[  m_{[G-k-1]}
-   c_{[G-k-1]}^2  s_{[G-k-1]}^2     \left(1+ m_{[G-k-1]} \right)^2
\right]
\label{cxyzpure}
\end{eqnarray}


\subsection{ Energy of the homogeneous Tree-Tensor-State and optimization of its $G$ parameters   }

For the pure Dyson model of Eqs \ref{dysonlca} and Eq. \ref{jpowertree},
where the magnetization $m_{[G]}$ depends only on the generation 
and where the correlation depends only on the generation $G$ and on the distance $r_k=2^k$,
the energy of the Tree-Tensor-State of Eq. \ref{edysonlca} becomes
\begin{eqnarray}
{\cal E}^{[G]}= - 2^G h m_{[G]}
- 2^{G-1}   \sum_{k=0}^{G-1} 
\sum_{a=x,y,z} \frac{ J^{a} }{  2^{k\omega_a}  } C^{a}_{[G]} (r_k=2^k)  
\label{edysonlcapure}
\end{eqnarray}
while the equivalent computation of the energy at any generation $g$ (Eq \ref{energieg}) reads
similarly 
\begin{eqnarray}
{\cal E}^{[g]} 
=E^{[g]} - 2^g h_{[g]} m_{[g]}
-  2^{g-1}   \sum_{k=0}^{g-1} 
\sum_{a=x,y,z} \frac{ J^{a}_{[g]} }{  2^{k\omega_a}  } C^{a}_{[g]} (r_k=2^k) 
\label{energiegpure}
\end{eqnarray}
in terms of the parameters $(E^{[g]},h_{[g]},J^{a}_{[g]})$ of the renormalized Hamiltonian.

The dependence with respect to the only parameter of the generation $g$
can be obtained from the energy computed at generation $(g+1)$
when the magnetization and the correlations of generation $(g+1)$ are written
in terms of the magnetizations and the correlations of generation $g$
via the recursions of Eqs \ref{recmupure}, \ref{Cpure1} and \ref{recorrepure}
\begin{eqnarray}
&& {\cal E}^{[g+1]} 
 =E^{[g+1]} - 2^{g+1} h_{[g+1]} m_{[g+1]}
-  2^{g}  \sum_{a=x,y,z} J^{a}_{[g+1]}  C^{a}_{[g+1]}(r_0=1) 
-  2^{g}   \sum_{k=0}^{g-1} 
\sum_{a=x,y,z} \frac{ J^{a}_{[g+1]} }{ 2^{\omega_a} 2^{k\omega_a}  } C^{a}_{[g+1]} (r_{k+1}=2^{k+1}) 
\label{energiegnextpure}
 \\
&& = E^{[g+1]} - 2^{g+1} h_{[g+1]}  c_{[g]}   s_{[g]}  (1+ m_{[g]})
-  2^{g} \left[ J^{x}_{[g+1]} \left(  c_{[g]}^2 -  s_{[g]}^2   m^{[g]}  \right)
+ J^{y}_{[g+1]} \left( s_{[g]}^2 -  c_{[g]}^2   m_{[g]}  \right)
+  J^{z}_{[g+1]} m_{[g]}
\right]
\nonumber \\
&&-  2^{g-1}   \sum_{k=0}^{g-1} 
\left[  \frac{2^{1-\omega_x} c_{[g]}^2 J^{x}_{[g+1]} }{  2^{k\omega_x}  } 
   C^{x}_{[g]}(r_k=2^k)
+  \frac{2^{1-\omega_y} s_{[g]}^2  J^{y}_{[g+1]} }{  2^{k\omega_y}  } 
    C^{y}_{[g]} (r_k=2^k)
 +\frac{2^{1-\omega_z} c_{[g]}^2 s_{[g]}^2 J^{z}_{[g+1]} }{  2^{k\omega_z}  } 
\left( 1+2 m_{[g]} + C^{z}_{[g]}(r_k=2^k) \right)
\right]
\nonumber
\end{eqnarray}

The identification with Eq. \ref{energieg} yields 
the renormalization rules for the couplings  (instead of Eq. \ref{rgj})
\begin{eqnarray}
J^{x}_{[g]}  && =2^{1-\omega_x} c_{[g]}^2 J^{x}_{[g+1]}
\nonumber \\
J^{y}_{[g]}  && =2^{1-\omega_y} s_{[g]}^2  J^{y}_{[g+1]}
\nonumber \\
J^{z}_{[g]}  && =2^{1-\omega_z} c_{[g]}^2 s_{[g]}^2 J^{z}_{[g+1]}
\label{rgjpure}
\end{eqnarray}
for the field (instead of Eq. \ref{rgh})
\begin{eqnarray}
   h_{[g]} 
 &&  =    2  c_{[g]}   s_{[g]} h_{[g+1]} 
-  s_{[g]}^2  J^{x}_{[g+1]} 
-  c_{[g]}^2 J^{y}_{[g+1]}
+  J^{z}_{[g+1]} \left[ 1 +2^{1-\omega_z} c_{[g]}^2 s_{[g]}^2
     \sum_{k=0}^{g-1} 
 \frac{ 1  }{  2^{k\omega_z}  } 
\right]
\label{rghpure}
\end{eqnarray}
and for the constant term (instead of Eq. \ref{rgconstant})
\begin{eqnarray}
  E^{[g]} 
&& = E^{[g+1]} - 2^{g} \left[ 2 c_{[g]}   s_{[g]}   h_{[g+1]}  
+   c_{[g]}^2   J^{x}_{[g+1]} 
+s_{[g]}^2  J^{y}_{[g+1]} 
+  2^{-\omega_z} c_{[g]}^2 s_{[g]}^2 J^{z}_{[g+1]}  \sum_{k=0}^{g-1} 
 \frac{ 1 }{  2^{k\omega_z}  } 
\right]
\label{rgepure}
\end{eqnarray}

Eq \ref{energiegnextpure} also gives the explicit dependence of the energy
with respect to the angle $\phi_{[g]}$ associated to the generation $g$
\begin{eqnarray}
 2^{-g} {\cal E}^{[g+1]} 
 && =2^{-g} E^{[g+1]} -  h_{[g+1]}  (1+ m_{[g]})\sin(2\phi_{[g]} )  -  J^{z}_{[g+1]} m_{[g]}
\nonumber \\
&&-    J^{x}_{[g+1]} \left( \frac{ (1- m_{[g]}) +(1+ m_{[g]})\cos(2\phi_{[g]} ) }{2}  \right)
- J^{y}_{[g+1]} \left( \frac{ (1- m_{[g]}) -  (1+ m_{[g]})\cos(2\phi_{[g]} ) }{2} \right)
\nonumber \\
&&-  2^{-1-\omega_x} [1+\cos(2\phi_{[g]} ) ]  J^{x}_{[g+1]}   \sum_{k=0}^{g-1} 
  \frac{ C^{x}_{[g]} (r_k=2^k) }{  2^{k\omega_x}  }  
-   2^{-1-\omega_y}  [1-\cos(2\phi_{[g]} ) ]  J^{y}_{[g+1]}
  \sum_{k=0}^{g-1}   \frac{  C^{y}_{[g]} (r_k=2^k) }{  2^{k\omega_y}  } 
\nonumber \\
&&-  2^{-2-\omega_z} \sin^2(2\phi_{[g]} ) J^{z}_{[g+1]}   \sum_{k=0}^{g-1} 
\left[ \frac{ \left( 1+2 m_{[g]} + C^{z}_{[g]} (r_k=2^k) \right)}{  2^{k\omega_z}  } 
\right]
\nonumber
\end{eqnarray}
The optimization equation with respect to the angle $\phi_{[g]}$ reads
\begin{eqnarray}
0= \frac{\partial ( 2^{-g}  {\cal E}^{[g+1]} ) }{ \partial \phi_{[g]}  } 
&& =   (1+ m_{[g]}) \left[ -  2 h_{[g+1]}  \cos (2\phi_{[g]} ) 
 +    (  J^{x}_{[g+1]} -   J^{y}_{[g+1]} ) \sin(2\phi_{[g]} ) \right]
\nonumber \\
&& +  \sin(2\phi_{[g]} ) 
\left[  2^{-\omega_x}   J^{x}_{[g+1]}   \sum_{k=0}^{g-1} 
  \frac{ C^{x}_{[g]} (r_k=2^k) }{  2^{k\omega_x}  }  
-    2^{-\omega_y}    J^{y}_{[g+1]}
  \sum_{k=0}^{g-1}   \frac{  C^{y}_{[g]}(r_k=2^k) }{  2^{k\omega_y}  } 
\right]
\nonumber \\
&&- \cos(2\phi_{[g]} ) \sin(2\phi_{[g]} ) 2^{-1-\omega_z}  J^{z}_{[g+1]}   \sum_{k=0}^{g-1} 
\left[ \frac{  1+2 m_{[g]} + C^{z}_{[g]} (r_k=2^k) }{  2^{k\omega_z}  } 
\right]
\label{optimipure}
\end{eqnarray}
If one neglects the contributions of the second and third lines, 
the first line allows to recover the usual criterion based on the diagonalization of the intra-Hamiltonian in each block
(Eq \ref{thetas}) for the angle (Eq \ref{phipure})
\begin{eqnarray}
2 \phi_{[g]} && =   \frac{\pi}{2}- 2 \theta^+_{[g]} 
\label{phipure2}
\end{eqnarray}


\section{ Scale-invariant Tree-Tensor-States for the critical pure Dyson models  }

\label{sec_criti}

In this section, we focus on the possible critical points of pure Dyson hierarchical quantum spin models,
where the corresponding homogeneous Tree-Tensor-State of the last section becomes in addition scale invariant.

\subsection{ Supplementary symmetry : scale invariance  }

At the critical points of the pure Dyson models discussed in the previous section,
the scale invariance means that the isometries do not even depend on the generation $g$ anymore,
so that the only remaining parameter is the angle $\theta^+ $ 
or the angle 
\begin{eqnarray}
\phi = \frac{\pi}{4}- \theta^{+} 
\label{phicriti}
\end{eqnarray}
so the parameters $c_{[g]}$ and $s_{[g]}$ of the previous section do not depend of $g$ anymore
\begin{eqnarray}
c && \equiv  \cos\left( \phi  \right) 
\nonumber \\
s && \equiv  \sin\left( \phi \right) 
\label{alphabetapurecriti}
\end{eqnarray}


\subsection{ Explicit solution for the one-point magnetization  }

The magnetization of Eq. \ref{solumupure} reduces to
\begin{eqnarray}
m_{[g]}
 =  \frac{ cs}{1-cs}  +  \left( cs \right)^g \left(  \frac{ 1-2 cs}{1-cs}  \right)
\label{solumugscale}
\end{eqnarray}
The dependence with respect to the generation $g$ comes only 
from the finite size and from the initial condition $m_{[0]}=+1$ at generation $g=0$.
In the thermodynamic limit $g\to +\infty$, the influence of this initial condition disappears
and the asymptotic magnetization is simply
\begin{eqnarray}
m_{[\infty]}  =  \frac{ cs}{1-cs} 
 = \frac{ \sin(2 \phi) }{2 - \sin(2 \phi)} 
\label{muinfty}
\end{eqnarray}


\subsection{ Explicit solutions for the two-point correlations  }

The two-point correlations of Eq. \ref{cxyzpure} simplify into
\begin{eqnarray}
C^{x}_{[G]}(r_k=2^k) 
&& =\left( c^2 \right)^{k} \left(  c^2 - s^2  m_{[G-k-1]} \right)
\nonumber \\
C^{y}_{[G]}(r_k=2^k) 
&& =\left( s^2 \right)^{k} \left(  s^2 - c^2  m_{[G-k-1]} \right)
\nonumber \\
 C^{z}_{[G]}(r_k=2^k) 
-   m_{[G]}^2 
&& =\left(   c^2 s^2 \right)^{k} \left[  m_{[G-k-1]}- (cs)^2 \left(1+m_{[G-k-1]} \right)^2  \right]
\label{cxyzpurecriti}
\end{eqnarray}

Again the dependence with respect to the generation $G$ 
comes only from the finite size via the magnetization $m_{[G-k-1]} $ of the last common ancestor.
In the thermodynamic limit $G \to +\infty$
where the asymptotic magnetization is given by Eq. \ref{muinfty},
the two-point-correlations become simple power-laws with respect to the distance $r_k=2^k$
\begin{eqnarray}
C^{x}_{[\infty]}(r_k=2^k) 
&& =\left( c^2 \right)^{k} \left(  c^2 - s^2  m_{[\infty]} \right) = \left( c^2 \right)^{k}  \frac{ c (c-s) }{1-cs} 
\equiv  \frac{ A_x }{ r_k^{2 \Delta_x} }
\nonumber \\
C^{x}_{[\infty]}(r_k=2^k) 
&& =\left( s^2 \right)^{k} \left(  s^2 - c^2  m_{[\infty]} \right)= \left( s^2 \right)^{k}  \frac{ s (s-c) }{1-cs} 
\equiv  \frac{ A_y }{ r_k^{2 \Delta_y} }
\nonumber \\
 C^{x}_{[\infty]}(r_k=2^k) 
-  m_{[\infty]}^2 
&& =\left(   c^2 s^2 \right)^{k} \left[  m_{[\infty]}- (cs)^2 \left(1+m_{[\infty]} \right)^2  \right]
= \left(   c^2 s^2 \right)^{k} \frac{ c s(1-2 cs)  }{(1-cs)^2} 
\equiv  \frac{ A_z }{ r_k^{2 \Delta_z} }
\label{cxyzpurecritithermo}
\end{eqnarray}
where the scaling dimensions $\Delta_a$ 
that govern the power-law decays 
with respect to the distance $r_k=2^k$ 
\begin{eqnarray}
  \Delta_x && = -\frac{ \ln \vert c \vert }{\ln 2}  = -\frac{ \ln \vert \cos\left( \phi  \right) \vert }{\ln 2}
\nonumber \\
  \Delta_y && = -\frac{ \ln \vert s \vert }{\ln 2} = -\frac{ \ln \vert \sin\left( \phi \right) \vert }{\ln 2}
\nonumber \\
  \Delta_z && = -\frac{ \ln \vert cs \vert }{\ln 2} =-\frac{ \ln \vert \cos\left( \phi  \right)\sin\left( \phi \right) \vert }{\ln 2} 
\label{delta}
\end{eqnarray}
and the amplitudes
\begin{eqnarray}
  A_x && = \frac{ c (c-s) }{1-cs} 
\nonumber \\
  A_y && =  \frac{ s (s-c) }{1-cs} 
\nonumber \\
  A_z && = \frac{ c s(1-2 cs)  }{(1-cs)^2} 
\label{ampli}
\end{eqnarray}
depend only on the angle $\phi$.

In summary, the scale-invariant Tree-Tenso-State depends on the single parameter $\phi$
that determines completely the magnetization $m_{[\infty]} $ and the power-law correlations $C^{x,y,z}_{[\infty]}(r_k=2^k)  $
in the thermodynamic limit $G \to +\infty$.
In the following subsections, we need to make the link with the properties of the Hamiltonian.


\subsection{ Scale-invariance of the renormalized Hamiltonian with the dynamical exponent $z$   }

The renormalization rules for the couplings (Eqs \ref{rgjpure}) become
\begin{eqnarray}
J^{x}_{[g]}  && =2^{1-\omega_x} c^2 J^{x}_{[g+1]} 
\nonumber \\
J^{y}_{[g]}  && =2^{1-\omega_y} s^2  J^{y}_{[g+1]} 
\nonumber \\
J^{z}_{[g]}  && =2^{1-\omega_z} (cs)^2 J^{z}_{[g+1]}
\label{rgjpurecriti}
\end{eqnarray}
while the renormalization rule for the field (Eq \ref{rghpure}) reads
(when the thermodynamic limit is taken in the last sum)
\begin{eqnarray}
   h_{[g]} 
 &&  =    2 c s  h_{[g+1]} -  s^2  J^{x}_{[g]} -  c^2 J^{y}_{[g]} 
+  J^{z}_{[g]} \left[ 1 +2^{1-\omega_z} (c s)^2     \sum_{k=0}^{\infty}  \frac{ 1  }{  2^{k\omega_z}  } \right]
\nonumber \\
&&  =    2 c s  h_{[g+1]} -  s^2  J^{x}_{[g]} -  c^2 J^{y}_{[g]} 
+  J^{z}_{[g]} \left[ 1 +     \frac{ 2 (c s)^2  }{   2^{\omega_z} -1  } \right]
\label{rghpurecriti}
\end{eqnarray}

At the critical point, the field and the couplings that do not vanish
 in the renormalized scale-invariant Hamiltonian
should all have the same scaling dimension given by the dynamical exponent $z$
\begin{eqnarray}
J^{a}_{[g]}  && \simeq 2^{-z} J^{a}_{[g+1]}
\nonumber \\
h_{[g]}  && \simeq 2^{-z} h_{[g+1]} 
\label{zdynamical}
\end{eqnarray}
As a consequence, the ratios $K^{a}_{[g]} \equiv \frac{ J^{a}{[g]} }{ h_{[g]}   } $
 associated to the couplings surviving in the scale-invariant renormalized Hamiltonian
should take fixed point values $K^a$ independent of the generation $g$
\begin{eqnarray}
K^{a}_{[g]} \equiv \frac{ J^{a}_{[g]} }{ h_{[g]}   }   = K^{a}_{[g+1]} \equiv K^a
\label{defratioka}
\end{eqnarray}

The optimization condition of Eq. \ref{optimipure} can be 
then rewritten in the thermodynamic limit $g \to +\infty$ as
\begin{eqnarray}
0
&& =   (1+ m_{[\infty]}) \left[ -  2   \cos (2\phi ) 
 +    (  K^x -   K^y ) \sin(2\phi ) \right]
 \nonumber \\ &&
 +  \sin(2\phi ) 
\left[  2^{-\omega_x}   K^x   \sum_{k=0}^{\infty}   \frac{ C^{x}_{[\infty]}(r_k=2^k) }{  2^{k\omega_x}  }  
-    2^{-\omega_y}    K^y  \sum_{k=0}^{\infty}   \frac{  C^{y}_{[\infty]}(r_k=2^k) }{  2^{k\omega_y}  } 
\right]
\nonumber \\
&&- \cos(2\phi ) \sin(2\phi ) 2^{-1-\omega_z} K^z \sum_{k=0}^{\infty} 
\left[ \frac{ \left( 1+ m_{[\infty]} \right)^2 + \left( C^{z}_{[\infty]}(r_k=2^k) - m_{[\infty]}^2 \right)}{  2^{k\omega_z}  } 
\right]
\label{optimipurecriti}
\end{eqnarray}
where the magnetization $m_{[\infty]} $ and the correlations $C^{a=x,y,z}_{[\infty]}(r_k=2^k) $
have been written in Eqs \ref{muinfty} and  \ref{cxyzpurecritithermo} in terms of the angle $\phi$.


\subsection{ Critical points of the pure Dyson quantum Ising model ($K^x \ne 0$ and $K^y=0=K^z$) }

Let us consider the possible critical points where the $y$-coupling and the $z$-coupling vanish
in the scale-invariant renormalized Hamiltonian $K^y=0=K^z$.
This will occur either if the $y$-couplings and the $z$-couplings already vanish in the initial condition $J^y=0=J^z$,
i.e. if the initial condition corresponds to the pure Dyson quantum Ising model,
or if their scaling dimensions in Eq. \ref{rgjpurecriti} make the two ratios $\frac{ J^{a=y,z}_{[g]} } { J^{x}_{[g]} }$
converge towards zero via renormalization.
Then the scale invariance with the dynamical exponent $z$ of Eq. \ref{zdynamical}
yields the two conditions from Eq. \ref{rgjpurecriti} and \ref{rghpurecriti}
\begin{eqnarray}
 2^{-z} && = \frac{ J^{x}_{[g]} }{  J^{x}_{[g+1]} }  = 2^{1-\omega_x} c^2 = 2^{-\omega_x} \left(1+\cos(2 \phi) \right)
\nonumber \\
 2^{-z} && = \frac{ h_{[g]} }{  h_{[g+1]} }    =    2 c s  -  s^2  K^x = \sin(2 \phi) - \frac{K^x}{2}\left(1-\cos(2 \phi) \right)
\label{zcritionlyx}
\end{eqnarray}
while the optimization condition of Eq. \ref{optimipurecriti}
becomes 
\begin{eqnarray}
0
 =   -  2   \cos (2\phi ) 
 +      K^x  \sin(2\phi ) \left[ 1
 +   \frac{ 2^{-\omega_x} }  {  (1+ m_{[\infty]}) }     \sum_{k=0}^{\infty} 
  \frac{ C^{x}_{[\infty]} (r_k=2^k) }{  2^{k\omega_x}  }  \right]
\label{optimipurecritix}
\end{eqnarray}

If one neglects the correlations $C^{x}_{[\infty]} (r_k=2^k) \to 0 $ in this optimization equation,
one recovers Eq. \ref{thetas} 
\begin{eqnarray}
0  =   -  2   \cos (2\phi )  +      K^x  \sin(2\phi ) 
\label{optimipurecritixintra}
\end{eqnarray}
based on diagonalization 
of the intra-Hamiltonian in each block (see Appendix \ref{app_diago}),
and the properties of the corresponding critical point have been discussed 
as a function of the power-law exponent $\omega_x$ in references \cite{c_dysontransverse,dyson_entang}
 (where $\omega_x$ is called $\sigma$).

When the correlations $C^{x}_{[\infty]} (r_k=2^k) $ are not neglected, 
Eq. \ref{optimipurecritix} reads using the explicit forms of Eqs \ref{muinfty} and \ref{cxyzpurecritithermo}
\begin{eqnarray}
0
&& =   -  2   \cos (2\phi ) 
 +      K^x  \sin(2\phi ) \left[ 1 +      \frac{ c (c-s) } { 2^{\omega_x} - c^2}
  \right]
 =  
    -  2   \cos (2\phi ) 
 +      K^x  \sin(2\phi )  \frac{ 2^{\omega_x}  -cs  } { 2^{\omega_x} - c^2} 
 \nonumber \\
&& =  
    -  2   \cos (2\phi ) 
 +      K^x  \sin(2\phi )  \frac{ 2^{1+\omega_x}  - \sin(2 \phi)  } { 2^{1+\omega_x} - \left(1+\cos(2 \phi) \right)}
 \label{optimipurecritixexpli}
\end{eqnarray}

In conclusion, for any decay exponent $\omega_x$ of the initial couplings,
the critical point describing the phase transition between the paramagnetic phase and the ferromagnetic phase
is characterized by its location $K^x$, by its dynamical exponent $z$ and by the angle $\phi$
that parametrizes the power-law correlation of Eq. \ref{cxyzpurecritithermo} :
these three parameters $(K^x,z,\phi)$ should be computed as a function of the exponent $\omega_x$ 
via the three equations : the two Eqs \ref{zcritionlyx} and Eq \ref{optimipurecritixexpli}.


\subsection{ Critical points where $K^x \ne 0$ and $K^y \ne 0$ while $K^z=0$  }

Let us now consider the case where both the $x$-coupling $K^x\ne 0$ and the $y$-coupling $K^y \ne 0$ survive in the renormalized scale invariant Hamiltonian,
while the $z$-coupling vanishes $K^z=0$.

Then the scale invariance with the dynamical exponent $z$ of Eq. \ref{zdynamical}
yields the following three conditions from Eq. \ref{rgjpurecriti} and \ref{rghpurecriti}
\begin{eqnarray}
 2^{-z} && = \frac{ J^{x}_{[g]} }{  J^{x}_{[g+1]} }  = 2^{1-\omega_x} c^2 = 2^{-\omega_x} \left(1+\cos(2 \phi) \right)
\nonumber \\
2^{-z} && = \frac{ J^{y}_{[g]} }{  J^{y}_{[g+1]} }  = 2^{1-\omega_y }  s^2 =2^{-\omega_y} \left(1- \cos(2 \phi) \right)
\nonumber \\
 2^{-z} && = \frac{ h_{[g]} }{  h_{[g+1]} }    =    2 c s  -  s^2  K^x-  c^2  K^y
 =  \sin(2 \phi) - K^x \frac{1- \cos(2 \phi)}{2}- K^y \frac{1+ \cos(2 \phi)}{2}
\label{zcritionlyxy}
\end{eqnarray}
while the optimization condition of Eq. \ref{optimipurecriti}
reads
\begin{eqnarray}
 2   \cos (2\phi )  -    (  K^x -   K^y ) \sin(2\phi )
=  \frac{ \sin(2\phi ) }{ 1+ m_{[\infty]}} 
\left[  2^{-\omega_x}   K^x   \sum_{k=0}^{\infty} 
  \frac{ C^{x}_{[\infty]} (r_k=2^k) }{  2^{k\omega_x}  }  
-    2^{-\omega_y}    K^y
  \sum_{k=0}^{\infty}   \frac{  C^{y}_{[\infty]} (r_k=2^k) }{  2^{k\omega_y}  } 
\right]
\label{optimipurecritixy}
\end{eqnarray}

If one neglects the correlations $C^{x,y}_{[\infty]} (r_k=2^k) \to 0 $ in this optimization equation,
one recovers Eq. \ref{thetas} 
\begin{eqnarray}
  2   \cos (2\phi )  -       (  K^x -   K^y )  \sin(2\phi ) =0
\label{optimipurecritixintra}
\end{eqnarray}
based on diagonalization 
of the intra-Hamiltonian in each block (see Appendix \ref{app_diago}).

When the correlations $C^{x,y}_{[\infty]} (r_k=2^k) $ are not neglected, 
Eq. \ref{optimipurecritixy} reads using the explicit forms of Eqs \ref{muinfty} and \ref{cxyzpurecritithermo}
\begin{eqnarray}
  2   \frac{ \cos (2\phi ) }{\sin(2\phi ) } = 
      K^x \frac{2^{\omega_x} - cs }{2^{\omega_x} - c^2}
 -    K^y \frac{2^{\omega_y} - cs }{2^{\omega_y} - s^2}
\label{optimipurecritixyexplici}
\end{eqnarray}

In conclusion, for decay exponents $(\omega_x,\omega_y)$ of the initial couplings,
the critical point is characterized by its location $(K^x,K^y)$, by its dynamical exponent $z$ and by the angle $\phi$
that parametrizes the power-law correlations of Eq. \ref{cxyzpurecritithermo} :
these four parameters $(K^x,z,\phi)$ should be computed as a function of the exponents  $(\omega_x,\omega_y)$
as follows :

(i) The ratio of the two first equations of Eq. \ref{zcritionlyx} yields that
 the angle $\phi$ is fixed by the difference between the exponents $\omega_x$ and $\omega_y$
\begin{eqnarray}
\tan^2(\phi) = \frac{s^2}{c^2}  = 2^{\omega_y-\omega_x} 
\label{tan2phi}
\end{eqnarray}

(ii) Then the dynamical exponent $z$ is fixed by the two first equations of Eq. \ref{zcritionlyxy}.

(iii) Then the fixed-point locations 
are given by the third equation of  \ref{zcritionlyxy} and the optimization Eq. \ref{optimipurecritixy}.


\section{ Conclusions }

\label{sec_conclusion}

In this paper, we have analyzed the simplest Tree-Tensor-States (TTS)
respecting the Parity and the Time-Reversal symmetries 
in order to describe the ground states of long-ranged quantum spin chains with or without disorder. 

We have first focused on inhomogeneous TTS for disordered long-ranged spin-chains.
Explicit formulas have been given for the one-point and two-point reduced density matrices 
as parametrized by the magnetizations and the two-point correlations.
We have then analyzed how the total energy of the TTS depends on each parameter of the TTS
in order to obtain the optimization equations and to compare them 
with the traditional block-spin renormalization procedure based on the diagonalization of some intra-block renormalized Hamiltonian.

We have then considered the pure long-ranged spin-chains in order to include the supplementary symmetries
in the TTS description, both for the off-critical region where the homogeneous TTS is made of isometries
that only depend on the generation, and for critical points where the homogeneous TTS becomes scale invariant 
with isometries that do not depend on the generation anymore.

Further work is needed to investigate whether the variational optimization with respect to parameters
can be also written explicitly for other types of tensor-states based on different entanglement architectures.


\appendix

\section{ Comparison with the isometries determined by the intra-block Hamiltonians }

\label{app_diago}

In this Appendix, we recall the usual block-spin RG rules based on the diagonalization 
of the intra-Hamiltonian in each block in order to compare with the variational optimization
of the isometries discussed in the text.
The renormalized intra-Hamiltonian associated to the block of the two children $(I1,I2)$ 
having the same ancestor $I$  reads
\begin{eqnarray}
H^{intra}_{I1,I2} && = 
- h_{I1} \sigma_{I1}^{z} 
- h_{I2} \sigma_{I2}^{z} 
- J^{z}_{I1,I2} \sigma_{I1}^{z}  \sigma_{I2}^{z} 
- J^{x}_{I1,I2} \sigma_{I1}^{x}  \sigma_{I2}^{x} 
-  J^{y}_{I1,I2} \sigma_{I1}^{y}  \sigma_{I2}^{y} 
\nonumber \\
\label{h12}
\end{eqnarray}

\subsection { Diagonalization in the parity sector $P=\sigma_{I1}^{z}  \sigma_{I2}^{z}=+$ }

In the parity sector $\sigma_{I1}^{z}  \sigma_{I2}^{z} =+$, the diagonalization of the Hamiltonian of Eq. \ref{h12}
\begin{eqnarray}
H^{intra}_{I1,I2} \ket{\uparrow \uparrow} && = - (h_{I1}+h_{I2} 
+ J^{z}_{I1,I2} ) \ket{\uparrow \uparrow} - ( J^{x}_{I1,I2}-J^{y} _{I1,I2} ) \ket{\downarrow \downarrow }
\nonumber \\
H^{intra}_{I1,I2} \ket{\downarrow \downarrow } && = - ( J^{x}_{I1,I2}-J^{y} _{I1,I2} ) \ket{\uparrow \uparrow}
 + (h_{I1}+h_{I2} - J^{z}_{I1,I2} ) \ket{\downarrow \downarrow }
\label{huvs}
\end{eqnarray}
leads to the two eigenvalues
\begin{eqnarray}
e^{ [P=+] }_{\pm} =  -  J^{z}_{I1,I2} \pm  \sqrt{(h_{I1}+h_{I2}  )^2 +( J^{x}_{I1,I2}-J^{y} _{I1,I2} )^2 }
\label{ener}
\end{eqnarray}
The eigenvector associated to the lowest eigenvalue $ e^{ [P=+] }_{-}$
is the kept state $\ket{ \psi_{I1,I2}^{+}  } $ of Eq. \ref{psip}
\begin{eqnarray}
\ket{ \psi_{I}^{+}  }
&&
= \cos (\theta^{+}_I) \ 
\ket{\uparrow \uparrow}
 + \sin (\theta^{+}_I) \ 
\ket{\downarrow \downarrow }
\label{psipa}
\end{eqnarray}
where the angle $\theta^{+}_I $ is fixed by the parameters of the renormalized intra-Hamiltonian of Eq. \ref{h12}
\begin{eqnarray}
\cos  ( 2 \theta^{+}_I)  && =  \frac{h_{I1}+h_{I2} }{ \sqrt{ (h_{I1}+h_{I2}  )^2 +( J^{x}_{I1,I2}-J^{y} _{I1,I2} )^2   }} 
\nonumber \\
\sin (2 \theta^{+}_I)  && = \frac{ J^{x}_{I1,I2}-J^{y} _{I1,I2}   }{ \sqrt{ (h_{I1}+h_{I2}  )^2 +( J^{x}_{I1,I2}-J^{y} _{I1,I2} )^2   }} 
\label{thetas}
\end{eqnarray}


\subsection { Diagonalization in the parity sector $P=\sigma_{I1}^{z}  \sigma_{I2}^{z}=-$ }

In the parity sector $\sigma_{I1}^{z}  \sigma_{I2}^{z}=-$, the diagonalization of the Hamiltonian of Eq. \ref{h12}
\begin{eqnarray}
H^{intra}_{I1,I2} \ket{\uparrow \downarrow} && =  (-h_{I1}+h_{I2} 
+ J^{z}_{I1,I2} ) \ket{\uparrow \downarrow} - ( J^{x}_{I1,I2}+J^{y} _{I1,I2} ) \ket{\downarrow \uparrow }
\nonumber \\
H^{intra}_{I1,I2} \ket{\downarrow \uparrow } && = - ( J^{x}_{I1,I2}+J^{y} _{I1,I2} )\ket{\uparrow \downarrow}
 + (h_{I1}-h_{I2} + J^{z}_{I1,I2} ) \ket{\downarrow \uparrow }
\label{ha}
\end{eqnarray}
leads to the two eigenvalues
\begin{eqnarray}
e^{ [P=-] }_{\pm} =    J^{z}_{I1,I2} \pm  \sqrt{(h_{I1}-h_{I2}  )^2 +( J^{x}_{I1,I2}+J^{y} _{I1,I2} )^2 }
\label{ea}
\end{eqnarray}
The eigenvector associated to the lowest eigenvalue $ e^{ [P=-] }_{-}$
is the kept state $\ket{ \psi_{I1,I2}^{-}  } $ of Eq. \ref{psim}
\begin{eqnarray}
\ket{ \psi_I^{-}  }
&&
= \cos (\theta^{-}_I) \ 
\ket{\uparrow \downarrow}
 + \sin (\theta^{-}_I) \ 
\ket{\downarrow \uparrow }
\label{psima}
\end{eqnarray}
where the angle $\theta^{-}_I $ is fixed by the parameters of the renormalized intra-Hamiltonian of Eq. \ref{h12}
\begin{eqnarray}
\cos  ( 2 \theta^{-}_I)  && =  \frac{h_{I1}-h_{I2} }{ \sqrt{ (h_{I1}-h_{I2}  )^2 +( J^{x}_{I1,I2}+J^{y} _{I1,I2} )^2   }} 
\nonumber \\
\sin (2 \theta^{-}_I)  && = \frac{ J^{x}_{I1,I2}+J^{y} _{I1,I2}   }{ \sqrt{ (h_{I1}-h_{I2}  )^2 +( J^{x}_{I1,I2}+J^{y} _{I1,I2} )^2   }} 
\label{thetaa}
\end{eqnarray}

\end{document}